\documentclass[11pt,twoside]{article}

\usepackage[usenames,x11names,table]{xcolor}
\usepackage[font={rm,small,sl}]{caption}
\usepackage{graphicx}
\usepackage{cancel}
\usepackage[obeyspaces]{url}% http://ctan.org/pkg/url
\usepackage[margin=1.1in]{geometry}
\usepackage{enumerate}
\usepackage{tikz}
\usepackage{amsmath}
\usepackage{amssymb}
\usepackage{subfigure}
\definecolor{darkgray}{rgb}{0.6,0.6,0.6}
\definecolor{darkgreen}{rgb}{0.3,0.5,0.3}
\definecolor{mygreen}{rgb}{0.0,0.4,0.0}
\definecolor{mygray}{rgb}{0.5,0.5,0.5}
\definecolor{mymauve}{rgb}{0.58,0,0.82}

\newcommand*\circled[1]{\tikz[baseline=(char.base)]{
            \node[shape=circle,draw,inner sep=0.6pt] (char) {#1};}}
\usepackage{listings}

\lstdefinestyle{MYC}{ 			   
  language=C,                      % the language of the code
  backgroundcolor=\color{white},   % choose the background color; you must add \usepackage{color} or \usepackage{xcolor}
  basicstyle=\scriptsize\ttfamily,        % the size of the fonts that are used for the code
  breakatwhitespace=false,         % sets if automatic breaks should only happen at whitespace
  breaklines=true,                 % sets automatic line breaking
  captionpos=b,                    % sets the caption-position to bottom
  commentstyle=\color{mygreen},    % comment style
  deletekeywords={...},            % if you want to delete keywords from the given language
  escapeinside={\%*}{*)},          % if you want to add LaTeX within your code
  extendedchars=true,              % lets you use non-ASCII characters; for 8-bits encodings only, does not work with UTF-8
  keepspaces=true,                 % keeps spaces in text, useful for keeping indentation of code (possibly needs columns=flexible)
  keywordstyle=\color{blue},       % keyword style
  otherkeywords={foreach, func, not},           % if you want to add more keywords to the set
  numbers=left,                    % where to put the line-numbers; possible values are (none, left, right)
  numbersep=10pt,                   % how far the line-numbers are from the code
  numberstyle=\tiny\color{black}, % the style that is used for the line-numbers
  rulecolor=\color{black},         % if not set, the frame-color may be changed on line-breaks within not-black text (e.g. comments (green here))
  showspaces=false,                % show spaces everywhere adding particular underscores; it overrides 'showstringspaces'
  showstringspaces=false,          % underline spaces within strings only
  showtabs=false,                  % show tabs within strings adding particular underscores
  stepnumber=1,                    % the step between two line-numbers. If it's 1, each line will be numbered
  stringstyle=\color{mymauve},     % string literal style
  tabsize=2,	                   % sets default tabsize to 2 spaces
  title=\lstname                   % show the filename of files included with \lstinputlisting; also try caption instead of title
}

\colorlet{color}{cyan!50}

\newcolumntype{L}[1]{>{\raggedright\let\newline\\\arraybackslash\hspace{0pt}}m{#1}}
\newcolumntype{C}[1]{>{\centering\let\newline\\\arraybackslash\hspace{0pt}}m{#1}}
\newcolumntype{R}[1]{>{\raggedleft\let\newline\\\arraybackslash\hspace{0pt}}m{#1}}

\newcommand{\smalltt}[1]{{\texttt{\small #1}}}

\begin{document}

% ****************** TITLE ****************************************

\title{How to Databasify a Blockchain: the Case of Hyperledger Fabric}

% ****************** AUTHORS **************************************

\author{Ankur Sharma\qquad Felix Martin Schuhknecht\qquad Divya Agrawal\qquad Jens Dittrich\\
\\
Big Data Analytics Group\\
Saarland Informatics Campus\\
\url{https://bigdata.uni-saarland.de}}

\date{\today}

\maketitle

\begin{abstract}

Within the last few years, a countless number of blockchain systems have emerged on the market, each one claiming to revolutionize the way of distributed transaction processing in one way or the other. Many blockchain features, such as byzantine fault tolerance (BFT), are indeed valuable additions in modern environments. However, despite all the hype around the technology, many of the challenges that blockchain systems have to face are fundamental transaction management problems. These are largely shared with traditional database systems, which have been around for decades already. 

These similarities become especially visible for systems, that blur the lines between blockchain systems and classical database systems. A great example of this is Hyperledger Fabric, an open-source permissioned blockchain system under development by IBM. By having a relaxed view on BFT, the transaction pipeline of Fabric highly resembles the workflow of classical distributed databases systems. 

This raises two questions: (1)~Which conceptual similarities and differences do actually exist between a system such as Fabric and a classical distributed database system? (2)~Is it possible to improve on the performance of Fabric by transitioning technology from the database world to blockchains and thus blurring the lines between these two types of systems even further?  
To tackle these questions, we first explore Fabric from the perspective of database research, where we observe  weaknesses in the transaction pipeline. We then solve these issues by transitioning well-understood database concepts to Fabric, namely transaction reordering as well as early transaction abort. Our experimental evaluation shows that our improved version Fabric++ significantly increases the throughput of successful transactions over the vanilla version by up to a factor of~$3$x. 

\end{abstract}

\section{Introduction}
\label{sec:introduction}

Blockchains are one of the hottest topics in modern distributed transaction processing. However, from the perspective of database research, one could raise the question: what makes these systems so special over classical distributed databases, that have been out there for a long time already?

The answer lies in byzantine fault tolerance: while classical distributed database systems require a trusted set of participants, blockchain systems are able to deal with a certain amount of \textit{maliciously} behaving nodes. 
This feature opens lots of new application fields such as transactions between organizations, that do not fully trust each other.
Unfortunately, ensuring BFT over all nodes of the network also heavily complicates transaction processing. If any node of the network is considered to be potentially malicious, a complex \textit{consensus mechanism} is required to ensure the integrity of the system. This consensus mechanism assures, that a transaction can only commit, if a majority of the network agrees to it.

Of course, the expensiveness of the consensus mechanism has also been observed by the blockchain engineers. Therefore, some systems trade BFT with performance by simply assuming certain parties of the network to be trustworthy. A great example for this is Hyperledger Fabric~\cite{fabric}, a popular open-source blockchain system introduced by IBM. In terms of BFT, it differs from other major players such as Bitcoin or Ethereum in two additional assumptions: First, Fabric assumes that the \textit{ordering service}, which globally orders all transactions that go through the system, is trustworthy. Second, it allows for the forming of so called \textit{organizations}. Within an organization, it is assumed that all peers trust each other. 
These two assumptions heavily simplify transaction processing. First of all, no complex consensus mechanism, such as PBFT~\cite{pbft}, is necessary. Second, the trust within an organization allows for a distribution of work within it and enables parallelism, as not every peer in the organization has to execute every transaction. 

With this relaxed view on BFT in mind, can we actually still consider Fabric a true blockchain system? A trustworthy ordering service, which globally arranges and schedules transactions, is a component that is present in classical distributed database systems as well. Further, the concept of an organization, in which all peers trust each other, is also present in distributed databases in its extreme form: all peers belong to a single organization. Besides of that, other core requirements of transaction management, such as ensuring transaction isolation or managing the data in a store, are essentially present one-to-one in both blockchains and database systems. 

At the example of Fabric, it becomes obvious that conceptually the lines between blockchain systems and distributed database systems are rather blurry. We believe this blurriness should be seen as a chance for the database community: Due to all these conceptual similarities, it becomes possible to transition well-understood database technology to the world of blockchains, significantly enhancing this new technology. 

The question remains which similarities can be exploited to transition database technology to Fabric and by how much can we improve on the state-of-the-art? To tackle this problem, we perform the following steps in this work:

\begin{enumerate}
\item To have a basis for the discussion, we first inspect the transaction flow of Hyperledger Fabric in the latest version~1.2 from a conceptual perspective. Fabric will serve as our case-study for the rest of the paper on how to "databasify" a blockchain system (Section~\ref{sec:fabric}).
\item Based on the analysis of the transaction flow in Fabric, we then inspect its components, that show the highest resemblance with those of database systems. We identify weaknesses in the implementation of Fabric of these components and describe, how database technology can be utilized to counter them. (Section~\ref{sec:sidebyside}).
\item We transition database technology to the transaction pipeline of Fabric. Precisely, we first improve on  
\textit{the ordering of transactions}. By default, the system orders transactions arbitrarily after simulation, leading to unnecessary serialization conflicts. To counter this problem, we introduce an advanced \textit{transaction reordering mechanism}, which aims at reducing the number of serialization conflicts between transactions within a block. This mechanism significantly increases the number of valid transactions, that make it through the system and therefore the overall throughput (Section~\ref{ssec:transaction_reodering}).
\item Next, we advance the \textit{abort of transactions}. By default, Fabric checks whether a transaction is valid right before the commit. This late abort unnecessarily penalizes the system by processing transactions, that have no chance to commit. To tackle this issue, we introduce the concept of \textit{early abort} to various stages of the pipeline. We identify invalid transactions as early as possible and abort them, assuring that the pipeline is not throttled by transactions that have no chance to commit eventually. A requirement for this concept is a \textit{fine-grained concurrency control mechanism}, by which we extend Fabric as well (Section~\ref{ssec:early_abort}).
These modifications significantly extend the vanilla Fabric, turning it into what we call Fabric++.
\item We perform an extensive experimental evaluation of the optimizations of Fabric++ under a custom blockchain benchmark simulating an asset transfer scenario. In total, we evaluate the transactional throughput under $108$~different configurations of Fabric and Fabric++ and show that we are able to significantly boost the performance over the vanilla version. Additionally, we vary the number of channels and clients and show, that our optimizations also have a positive impact on the scaling capabilities of the system (Section~\ref{sec:ea}). 
\end{enumerate}

\section{Hyperledger Fabric}
\label{sec:fabric}

Before diving into the conceptual similarities and differences between Fabric and distributed database systems, we have to understand the workflow of Fabric. Let us describe in the following section how it  behaves in the latest version~1.2. 

\subsection{High-level Workflow}  

\begin{figure}[!htb]
    \begin{center}
        \includegraphics[width=12cm]{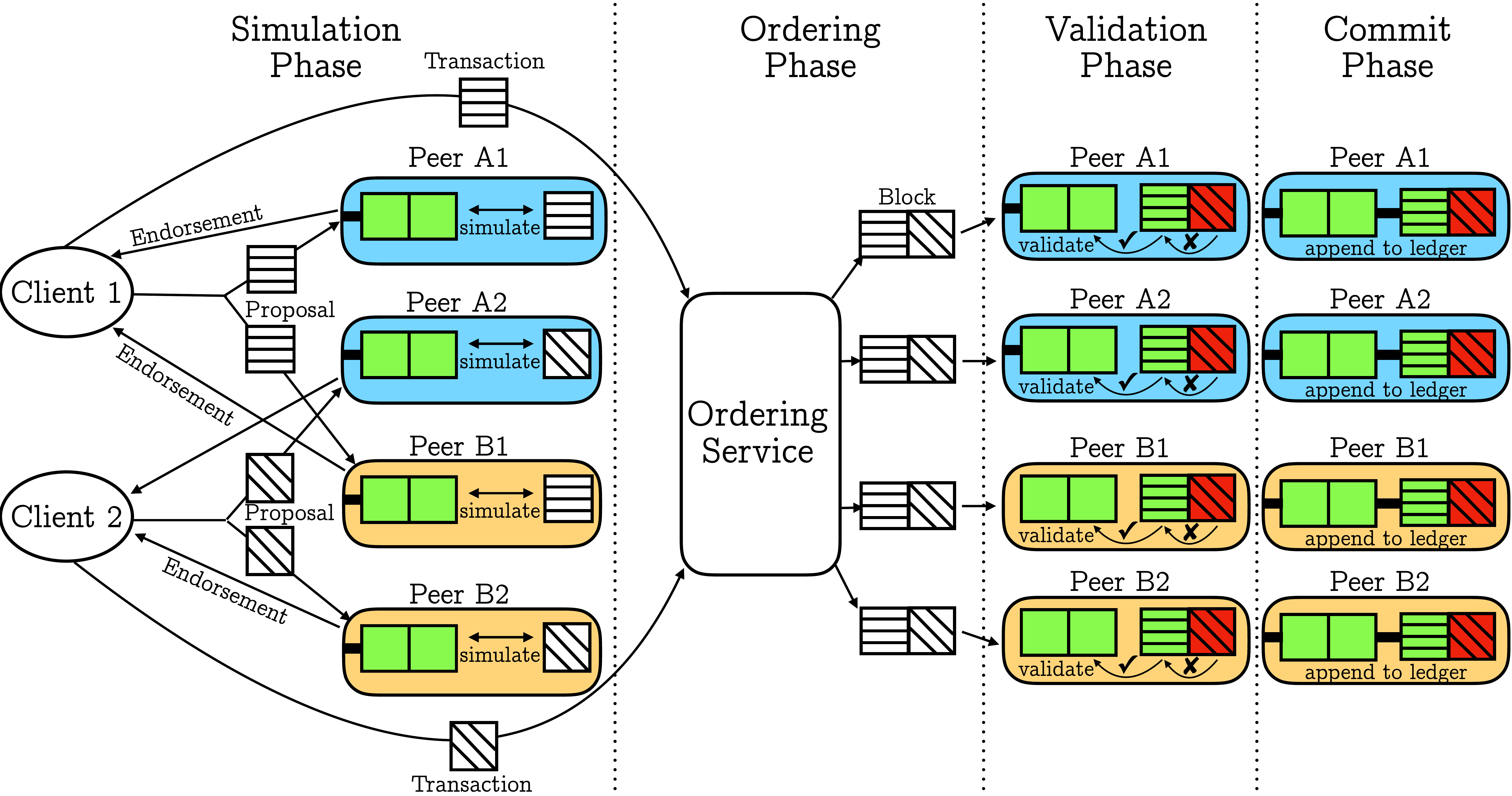}
    \end{center}
    \caption{High-level workflow of Fabric.}
    \label{figs:fabric_workflow_highlevel}
\end{figure}

At its core, Fabric follows a \textit{simulate-order-validate-commit} workflow, as shown in Figure~\ref{figs:fabric_workflow_highlevel}:
%\footnote{In the original paper~\cite{fabric}, this workflow is called \textit{execute-order-validate}. We change this in our description to avoid confusion. In particular, `execute' can easily be misunderstood to change the current state in the system --- which is not the case. In addition, the original `validate' phase contains both elements of `validate' and `commit'.}: 

\noindent (1)~In the \textbf{simulation phase}, a client submits a transaction proposal to a specified subset of the peers, called the endorsement peers or \textit{endorsers}. The endorsers simulate the effects of the transaction proposal against a local copy of their current state. Interestingly, none of the writes become durable in the current state at this point. If the endorsers endorse the transaction proposal, an actual transaction is formed from the execution result, that is then sent to the ordering service (via the client). 

\noindent (2)~In the \textbf{ordering phase}, the ordering service establishes a global order among all received transactions and distributes the ordered transactions at the granularity of blocks to all peers of the network. 

\noindent (3)~In the \textbf{validation phase}, all peers individually validate the transactions within the received blocks in terms of endorsement policy and serializability.

\noindent (4)~In the \textbf{commit phase}, the blocks are  appended to the local ledger and the changes made by the valid transactions are applied to the current state.

Following these four phases assures that each honest peer stores the same transaction sequence.

\subsection{Architecture}

Fabric is a \textit{permissioned} blockchain system, meaning all peers of the network are known at any point in time. Peers are grouped into \textit{organizations}, which typically host them. For example, two companies trading with each other could each host an organization of $10$~machines, forming a network of $20$~peers. Within an organization, all peers trust each other.  

Each peer runs a local instance of Fabric. This instance includes a copy of the \textit{ledger}, containing the ordered sequence of all transactions that went through all four phases. This includes both valid and invalid transactions. Apart from the ledger, each peer also contains the \textit{current state} in form of a state database. The current state can be seen as an optimization of the ledger: while the ledger simply contains the sequence of all processed transactions, the current state represents the state after the \textit{application} of all \textit{valid} transactions in the ledger to the initial state. Fabric implements the current state in form of a versioned key-value-store. For every key in the store, a pair of value and version-number is kept, where the version-number\footnote{The version number is actually composed of transaction-ID and the block-ID, see Section~\ref{ssec:early_abort_simulation} for details.} counts the number of changes that already happened to the value of this key.

Apart from the peers, which play an important role both in the simulation phase and the validation phase, there is a separate instance called the \textit{ordering service}, which is the core component of the ordering phase and assumed to be trustworthy.  Although it can be composed out of multiple machines for fault tolerance, it is a central service responsible for establishing a global order among all transactions.

\subsection{Running Example}

With the basic components of the architecture in mind, let us now discuss how transactions flow through the system. To do so, we present a simple running example in Figure~\ref{figs:fabric_workflow_execute} (simulation phase), Figure~\ref{figs:fabric_workflow_order} (ordering phase), and Figure~\ref{figs:fabric_workflow_validate} (validation and commit phase), where two organizations $A$ and $B$ want to transfer money between each other. 

Each organization contributes two peers to the network. The balances of the organizations are stored by two variables $BalA$ and $BalB$, where $BalA$ stores the value~$100$ in its current version~$v_3$ and $BalB$ stores~$50$ in version~$v_2$.
We can also see that the ledger already contains six transactions~$T_1$ to~$T_6$, where the four transactions~$T_1$, $T_2$, $T_4$, and $T_6$ were valid ones and lead to the current state. The transactions $T_3$ and $T_5$ were invalid transactions. They are still stored in the ledger, although they did not pass the validation phase.

\subsection{Simulation Phase}
\label{ssec:simulation}

\begin{figure}[!htb]
    \begin{center}
        \includegraphics[width=12cm]{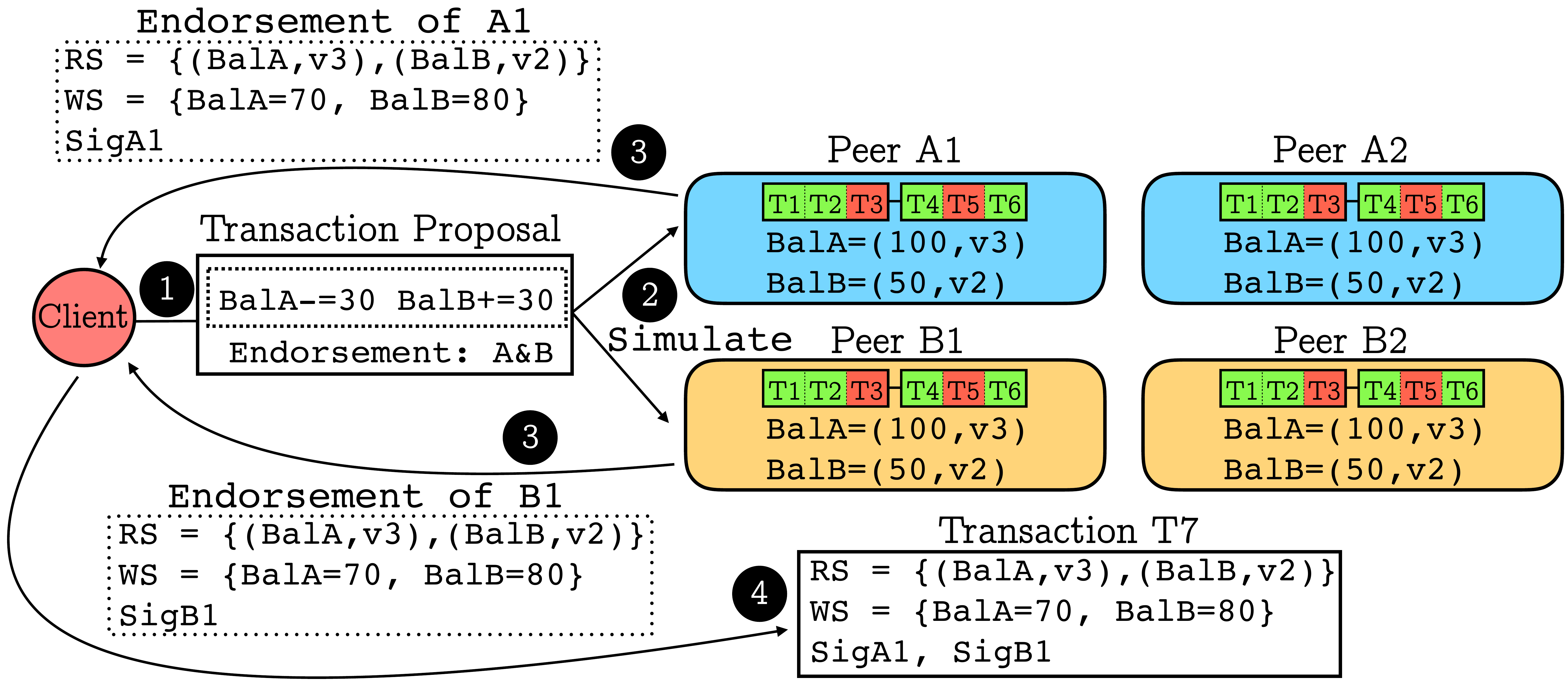}
    \end{center}
    \caption{Simulation phase.}
    \label{figs:fabric_workflow_execute}
\end{figure}

Transaction processing starts with the \textit{simulation} phase in Figure~\ref{figs:fabric_workflow_execute}. In \textbf{step}~\circled{1}, a client proposes a \textit{transaction proposal} (or short \textit{proposal}) to the system. In our example, the  proposal intends to transfer the amount of~$30$ from $BalA$ to $BalB$. The two involved operations \mbox{$BalA$-=$30$} and \mbox{$BalB$+=$30$} are expressed in a smart contract\footnote{Smart contracts are typically called \textit{chaincodes} in Fabric. However, as they do not conceptually differ from smart contracts in blockchain systems such as Ethereum, we stick to this term throughout the paper.}, an arbitrary program, that is bound to the proposal. 
Additionally to the smart contract, an endorsement policy must be specified. It determines which and/or how many peers have to endorse the proposal. In our example of money transfer between two organizations, a reasonable endorsement policy is to request endorsement from one peer of each organization --- like two lawyers, preserving and defending the individual rights of their clients.  

Therefore, in \textbf{step}~\circled{2}, the proposal is sent to the two endorsement peers~$A1$ and~$B1$ according to the policy. These two peers now individually simulate the smart contract (\mbox{$BalA$-=$30$}, \mbox{$BalB$+=$30$}), that is bound to the proposal, against their local current state. 
Note that, as the name suggests, the simulation of the smart contract against the current state does \textit{not change} the current state in any way. Instead, each endorsement peer builds an auxiliary read set~$RS$ and a write set~$WS$ during the simulation to keep track of all accesses that happen.
In our case of money transfer over the amount of~$30$, the smart contract first reads the two current balances~$BalA$ and~$BalB$ along with their current version-numbers. Second, the smart contract updates the two balances according to the transferred amount, resulting in the new balances~$BalA=70$ and $BalB=80$.
Overall, this builds the following read and write set:
$$RS=\{(BalA, v_3), (BalB, v_2)\} \hspace{0.2cm}WS=\{ BalA=70, BalB=80 \}$$

\noindent In this sense, the simulation of the smart contract is actually only a monitoring of the execution effects. The reason for performing only a simulation is that in this phase, we can not be sure yet whether this transaction will be allowed to commit eventually -- this check will be performed later in the validation phase. 

After the simulation of the smart contract on all endorsement peers, in \textbf{step}~\circled{3}, the endorsement peers return their individually computed read and write sets to the client, that sent the transaction proposal. Additionally, they return a signature of their simulation, that will be relevant in the validation phase in Section~\ref{ssec:validation}.
If all read sets and write sets match\footnote{They might not match due to non-determinism in the smart contract or due to malicious behavior of the endorsement peer(s).}, in \textbf{step}~\circled{4}, the actual \textit{transaction} (called~$T_7$ in the following) is formed from the results of the endorsement. This transaction~$T_7$ now contains the effects of the execution in form of the read and write set as well as all signatures and can be passed on to the ordering service.

\subsection{Ordering Phase}
\label{ssec:ordering}

\begin{figure}[!htb]
    \begin{center}
        \includegraphics[width=12cm]{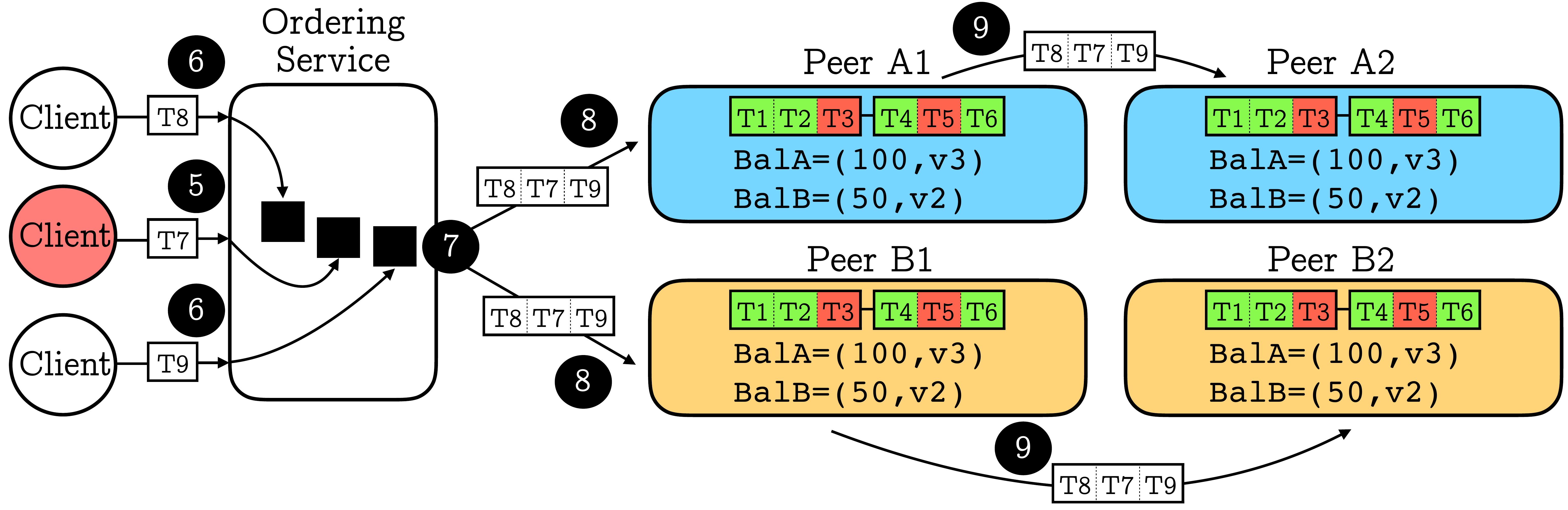}
    \end{center}
    \caption{Ordering phase.}
    \label{figs:fabric_workflow_order}
\end{figure}

As mentioned, the central component of the \textit{ordering} phase is the ordering service, that we visualize in Figure~\ref{figs:fabric_workflow_order}. It receives all transactions, that made it through the simulation phase. Consequently, it receives in \textbf{step}~\circled{5} our transaction~$T_7$, that we followed through the simulation phase in Section~\ref{ssec:simulation}. In \textbf{step}~\circled{6}, we assume that it also receives two other transactions~$T_8$ and~$T_9$, that were endorsed in parallel to~$T_7$. 

The ordering service now has the sole purpose of establishing a global order among the transactions. It treats the transactions in a black box fashion and does not inspect the transaction semantics, such as the read and write set, in any way. By default, it essentially arranges the transactions in the order in which they arrive, resulting in what we call for the rest of the paper the \textit{arrival order}. In \textbf{step}~\circled{7}, the ordering service now outputs the ordered stream of transactions in form of \textit{blocks}, containing a certain number of transactions. Outputting whole blocks instead of individual transactions reduces the pressure on the network, as less communication overhead is produced.

Finally, the generated block is distributed to \textit{all} four peers of the network to start the validation phase. Note that there is no guarantee that all peers receive a block at the same time, as the distribution happens partially from ordering service to peers directly as shown in~\textbf{step}~\circled{8} and partially between the peers using a gossip protocol as shown in~\textbf{step}~\circled{9}. 
However, the service assures that all peers receive the blocks in the same order.

\subsection{Validation and Commit Phase}
\label{ssec:validation}

\begin{figure}[!htb]
    \begin{center}
        \includegraphics[width=12cm]{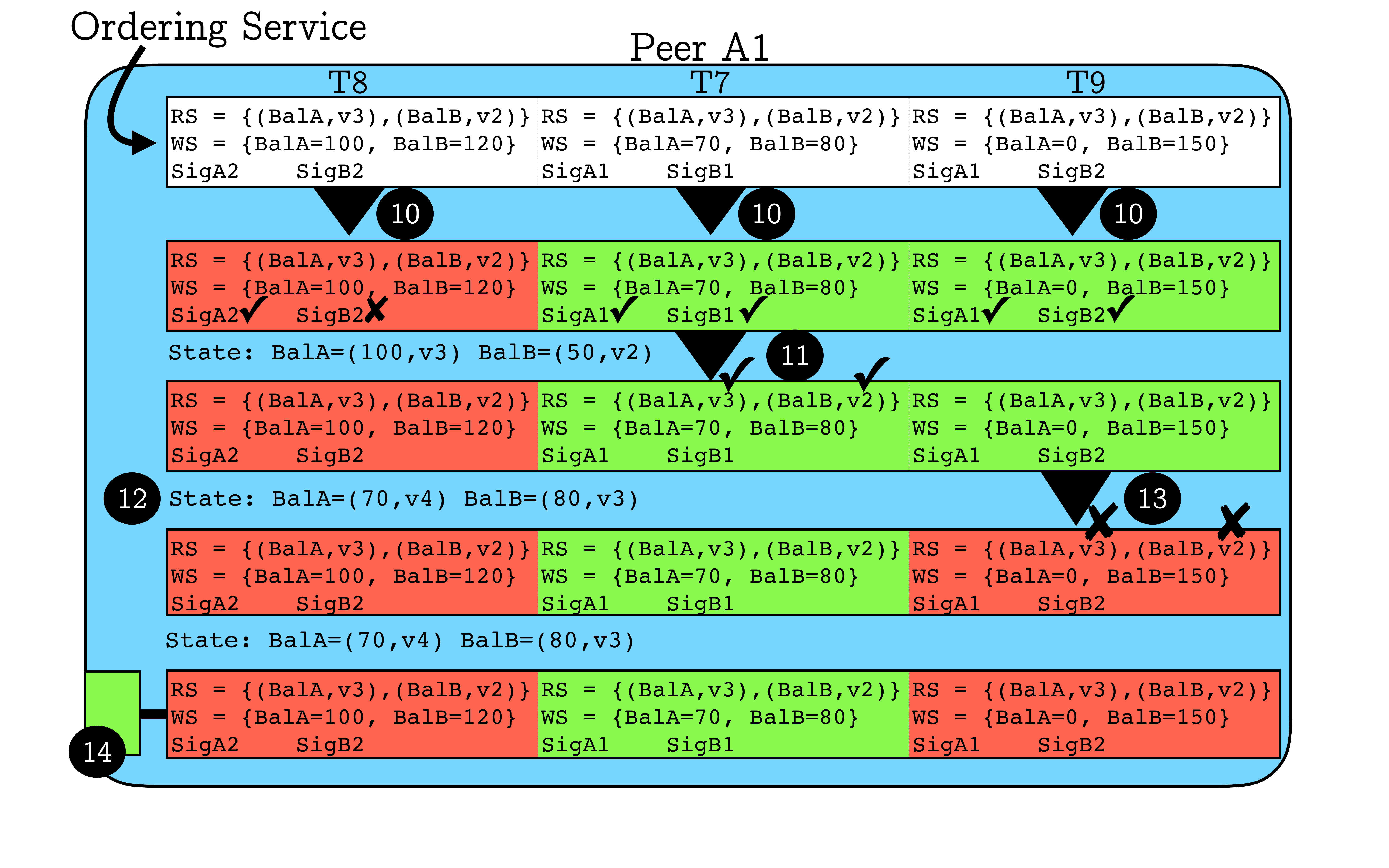}
    \end{center}
    \caption{Validation and Commit Phase.}
    \label{figs:fabric_workflow_validate}
\end{figure}

When a block arrives at a peer, the \textit{validation} phase starts, visualized in Figure~\ref{figs:fabric_workflow_validate} for peer~$A1$. The three remaining peers execute the same validation process. Overall, the validation phase has two purposes.

\subsubsection{Endorsement Policy Evaluation} 
\label{sssec:endorsement_policy_eval}

The first purpose is to validate the transactions in the block with respect to the \textit{endorsement policy}. For example, it is possible that a malicious transaction was generated by a malicious client and a malicious peer in conspiracy to take advantage of the money transfer. Let us assume that transaction~$T_8$ is such a malicious transaction and that the malicious client, which proposed~$T_8$, works together with peer~$A2$, which is also malicious. 
Instead of using the legit write set~$WS_{B2}=\{BalA=30, BalB=120\}$ from B2, the client creates a proposal with the write set~$WS_{A2}=\{BalA=100, BalB=120\}$, that it received from its collaborator A2.

How is this transaction~$T_8$ now detected in the validation phase? The key to this lies in the signatures~$Sig_{A2}$ and $Sig_{B2}$, that the endorsement peers generate at the end of the simulation phase. The signature is computed over the read and write set, the executed smart contract, and the used endorsement policy. The client receives these cryptographically secure signatures and must pack them into the transaction along with the read and write set. The peers that validate the transaction recompute the signatures of all endorsement peers, that were responsible for transaction~$T_8$ and compare the signatures with the received ones~$Sig_{A2}$ and $Sig_{B2}$. 
	In our example, in \textbf{step}~\circled{10}, the peers detect that the signature of the honest peer~$Sig_{B2}$ does not match to the one they computed from the received write set and thus, would classify~$T_8$ as invalid. $T_7$ and $T_9$, the remaining transactions in the block, are evaluated in parallel. Their signatures match the ones computed from the read and write set and therefore, these transactions are valid with respect to the endorsement policy. 

\subsubsection{Serializability Conflict Check}
\label{sssec:serializability_conflict_check}

The second purpose of the validation is to analyze the transactions with respect to \textit{serializability conflicts}, that can arise from the order of transactions. For every transaction, it must be checked whether the version-numbers of all keys in the read set match the version-numbers in the current state. Only if this is the case, a transaction operates on an up-to-date state. 
Considering our example, let us perform the serializability conflict check for the received block. $T_8$ is already marked as invalid as it did not pass the endorsement policy evaluation, so it is not checked again. $T_7$ passed the endorsement policy evaluation and is now tested for serialization conflicts in \textbf{step}~\circled{11}. Its read set is~$RS=\{(BalA, v_3), (BalB, v_2)\}$. The version numbers of $BalA$ and $BalB$ in the read set match the ones of the current state and therefore, $T_7$ is marked as valid. As a consequence, in \textbf{step}~\circled{12}, the write set of $T_7$, namely $WS=\{ BalA=70, BalB=80 \}$ is written to the current state. This changes the current state to $BalA=(70,v_4)$~and~$BalB=(80,v_3)$. Note that the version-numbers of the modified variables are incremented.  

The next transaction to be checked is~$T_9$ in \textbf{step}~\circled{13}. Let us assume it also performs a money transfer and has the following read and write set: 
$$ RS=\{(BalA, v_3), (BalB, v_2)\} \hspace{0.2cm} WS=\{ BalA=0, BalB=150 \} $$
This transaction will not pass the conflict check, as it read $BalA$ in version~$v_3$ and $BalB$ in version~$v_2$, while the current state already contains $BalA$ in version~$v_4$ and $BalB$ in version~$v_3$. Therefore, it operated on outdated data and is marked as invalid. As a consequence, its write set is not applied to the current state and simply discarded.

Finally, after validating all transactions of the block, in \textbf{step}~\circled{14} the entire block is appended to the ledger along with the information about which transactions are valid or invalid.

\section{Blurred Lines: Fabric vs Distributed Database Systems}
\label{sec:sidebyside}

As we now have an understanding of the workflow of Fabric, we are able to discuss its architecture in relation to distributed database systems. In particular, we are interested in aspects of Fabric, that are (a) conceptually shared with distributed database systems, but (b) have potential for the application of database technology.

\subsection{The Importance of Transaction Order}
\label{ssec:importance_of_order}
\vspace{-0.1cm}
The first component we look at is the ordering mechanism. Such a component is also present in any distributed database system with transaction semantics and therefore a great candidate for transitioning database technology to Fabric.

As described in Section~\ref{ssec:ordering}, Fabric relies on a single trustworthy ordering service for ordering transactions. Since Fabric simulates the smart contracts bound to proposals \textit{before} performing the ordering, the order actually has an influence on the number of serialization conflicts between transactions. Again, this is a property shared with any parallel database system, that separates transaction execution from transaction commit.   

In ordering transactions, various different strategies are possible: The simplest option is to arbitrarily order them, for instance in the order in which they arrive. While this arrival order is fast to establish, it can lead to serialization conflicts, that are \textit{potentially unnecessary}. These conflicts increase the number of invalid transactions, which must be resubmitted by the client. 
Unfortunately, the vanilla Fabric follows exactly this naive strategy. This is caused by the design decision that the ordering service is not supposed to inspect the transaction semantics, such as the read and write set, in any way. Instead, it simply leaves the transactions in the order in which they arrive. This strategy can be problematic, as the example in Table~\ref{table:naive_order} shows. 
In this example, four transactions are scheduled in the order in which they arrive, namely~$T_1 \Rightarrow T_2 \Rightarrow T_3 \Rightarrow T_4$, where $T_1$ updates the key~$k_1$ from version~$v_1$ to $v_2$. Since the transactions~$T_2$, $T_3$, and $T_4$ each read $k_1$ in version~$v_1$ during their simulations, they have no chance to commit, as they operated on an outdated version of the value of~$k_1$. They will be identified as invalid in the validation phase and the corresponding transaction proposals must be resubmitted by the client, resulting in a new round of simulation, ordering, and validation.

\begin{table}[!htb]
	\setlength{\tabcolsep}{3pt}
	\begin{center}
		\caption{For the order $T_1 \Rightarrow T_2 \Rightarrow T_3 \Rightarrow T_4$, only one out of four transactions is valid: $T_2$, $T_3$, and $T_4$ read the outdated version~$v_1$ of key~$k_1$, that has been updated by $T_1$ to $v_2$ before. }
		\begin{tabular}{ L{3cm} | R{3cm} | R{3cm} | R{3cm} }
			\hline
			\textbf{Transaction} & \textbf{Read Set} & \textbf{Write Set} & \textbf{Is Valid?}\\
			\hline
			% 50 cols, small pages
			%\hline
			1. $T_1$ & --- & \textcolor{red}{$(k_1, v_1 \rightarrow v_2)$} & \textcolor{blue}{\checkmark}  \\\hline
			2. $T_2$ & \textcolor{red}{$(k_1, v_1)$}$, (k_2, v_1)$ & $(k_2, v_1 \rightarrow v_2)$ & \textcolor{red}{$\times$}  \\\hline
			3. $T_3$ & \textcolor{red}{$(k_1, v_1)$}$, (k_3, v_1)$ & $(k_3, v_1 \rightarrow v_2)$ & \textcolor{red}{$\times$}  \\\hline
			4. $T_4$ & \textcolor{red}{$(k_1, v_1)$}$, (k_3, v_1)$ & $(k_4, v_1 \rightarrow v_2)$ & \textcolor{red}{$\times$}  \\
			\hline
		\end{tabular}
		\label{table:naive_order}
	\end{center}
\end{table}

Interestingly, for the four transactions from the previous example, there exists an order that is conflict free. In the schedule $T_4 \Rightarrow T_2 \Rightarrow T_3 \Rightarrow T_1$, as shown in Table~\ref{table:smart_order}, all four transactions are valid, as their read and write sets do not conflict with each other in this order. 

\begin{table}[!htb]
%\vspace{-0.3cm}
	\setlength{\tabcolsep}{3pt}
	\begin{center}
		\caption{The order $T_4 \Rightarrow T_2 \Rightarrow T_3 \Rightarrow T_1$ results in all four transactions being valid.}
		\begin{tabular}{ L{3cm} | R{3cm} | R{3cm} | R{3cm} }
			\hline
			\textbf{Transaction} & \textbf{Read Set} & \textbf{Write Set} & \textbf{Is Valid?}\\
			\hline
			% 50 cols, small pages
			%\hline
			1. $T_4$ & $(k_1, v_1)$$, (k_3, v_1)$ & $(k_4, v_1 \rightarrow v_2)$ & \textcolor{blue}{\checkmark}  \\\hline
			2. $T_2$ & $(k_1, v_1)$$, (k_2, v_1)$ & $(k_2, v_1 \rightarrow v_2)$ & \textcolor{blue}{\checkmark}  \\\hline
			3. $T_3$ & $(k_1, v_1)$$, (k_3, v_1)$ & $(k_3, v_1 \rightarrow v_2)$ & \textcolor{blue}{\checkmark}  \\\hline
			4. $T_1$ & --- & $(k_1, v_1 \rightarrow v_2)$ & \textcolor{blue}{\checkmark}  \\
			\hline
		\end{tabular}
		\label{table:smart_order}
	\end{center}
\end{table}

This example shows that the vanilla orderer of Fabric misses a chance of removing \textit{unnecessary} serialization conflicts.
While this problem is new to the blockchain domain, as blockchains typically offer only a serial execution of transactions, within the database community, this problem is actually well known. There exist reordering mechanisms which aim at minimizing the number of serialization conflicts via a reordering of transactions~\cite{reordering, reordering2, abort_dependency_detection}. However, in a database system, it is typically avoided to buffer a large number of incoming transactions before processing as low latency is mandatory. Thus, reordering is not always an option in such a setup. Fortunately, as blockchain systems buffer the incoming transactions anyways to group them into blocks, this gives us the opportunity to apply sophisticated transaction reordering mechanisms without introducing significant overhead. 

We will add such a transaction reordering mechanism to Fabric in Section~\ref{ssec:transaction_reodering}, which significantly enhances the number of valid transactions, that make it through the system.

\subsection{On the Lifetime of Transactions}
\vspace{-0.1cm}

The second aspect we look at from a database perspective tackles the lifetime of transactions within the pipeline. In Fabric, every transaction that goes through the system is either classified as valid or as invalid with respect to the validation criteria. In the vanilla version, this classification happens in the validation phase right before the commit phase. A severe downside of this form of \textit{late abort} is that a transaction, that violated the validation criteria already in an earlier phase, is still processed and distributed across all peers. This penalizes the whole system with unnecessary work, throttling the performance of valid transactions. Besides, this concept also delays the abort notification to the client. 

We have to distinguish in which phase a violation happens. First, a violation can occur already in the simulation phase, in form of so called \textit{cross-block conflicts}, meaning a transaction from a later block, which is currently in the simulation phase, conflicts with a valid transaction from an earlier block. 
Second, a violation can occur as well as in the ordering phase, in form of \textit{within-block conflicts} between conflicting transactions in a single block. 

Let us look at these two scenarios in isolation in Section~\ref{sssec:early_abort_simulation} and Section~\ref{sssec:early_abort_ordering}, respectively. 
\subsubsection{Violation in the simulation phase (cross-block conflicts)}
\label{sssec:early_abort_simulation}
To understand the problem in the simulation phase, let us look at the following situation and how the vanilla version of Fabric handles it. 
Let us assume there are four transactions~$T_1$, $T_2$, $T_3$, and $T_4$ that are currently in the ordering phase and that end up in a block of size four, which is shipped to all peers for validation. Before the validation of that block starts within a peer~$P$, the smart contract of a transaction proposal~$T_5$ starts its simulation in~$P$. To do so, it acquires a read lock\footnote{The read lock can be shared by multiple simulation phases, as they do not modify the current state.} on the \textit{entire} current state. While the simulation is running, the block has to wait for the validation, as it has to acquire an \textit{exclusive} write lock on the current state. 
The problem in this situation is: if $T_1$, $T_2$, $T_3$, or $T_4$ write the value of a key, that is read by $T_5$, then $T_5$ simulates on stale data. Therefore, in the moment of the read, the transaction becomes virtually invalid. Still, in the vanilla version of Fabric, this stale read is not detected before the validation phase of $T_5$. Thus, $T_5$ would continue its simulation and go through the ordering phase, just to be invalidated in the very end.

\subsubsection{Violation in the ordering phase (within-block conflicts)}
\label{sssec:early_abort_ordering}
Apart from conflicts across blocks, there can be conflicts between transactions within a block. These conflicts appear after putting the transactions into a particular order in the ordering phase. For instance, the example from Table~\ref{table:naive_order} in Section~\ref{ssec:importance_of_order} showed a schedule, where the three transactions~$T_2$, $T_3$, and $T_4$ individually conflict with the previously scheduled transaction~$T_1$ of the same block. Unfortunately, these conflicts are not detected within the orderer of the vanilla version of Fabric. The block containing $T_2$, $T_3$, and $T_4$ would be distributed across all peers of the network for validation, although $3/4$ of transactions within the block are virtually invalid. As before, this originates from the design decision that the ordering service does not inspect transaction semantics.

The mentioned situations show that Fabric misses several chances to abort transactions right at the time of violation. In contrast to that, database systems are typically very eager in aborting transactions~\cite{early_abort_paper}, as it decreases network traffic and saves computing resources. This concept of "cleaning" the pipeline as early as possible is called \textit{early abort} in the context of databases, which apply this concept in various flavors. For instance, besides of the early abort of transactions, that violate certain criteria, database systems eliminate records from the query result set as early as possible by pushing down selection and projection operations in the query plan.  

To overcome the mentioned problems, we will apply the concept of early abort at several stages of the transaction processing pipeline of Fabric. By this, we assure to utilize the available resources with meaningful work to the extend. We will detail this in Section~\ref{ssec:early_abort}.

\section{Fabric++}
\label{sec:databasify}

We have outlined the problems of Fabric and how they relate to key problems known in the context of database systems. Let us now see precisely how we counter them. First, in Section~\ref{ssec:transaction_reodering}, we introduce a transaction reordering mechanism, that aims at minimizing the number of unnecessary within-block conflicts. Second, in Section~\ref{ssec:early_abort}, we introduce early transaction abort to several stages of the Fabric pipeline. This also involves the introduction of a fine-grained concurrency control mechanism.

\subsection{Transaction Reordering}
\label{ssec:transaction_reodering}

When reordering a set of transactions~$S$, multiple challenges must be faced. First, we have to identify which transactions of~$S$ actually conflict with each other with respect to the actions they perform. Precisely, we have a conflict between two transactions~$T_i$ and $T_j$ (denoted as $T_i \nrightarrow T_j$), if $T_i$ writes to a key that is read by $T_j$. In this case, $T_i$ must be ordered \textit{after} $T_j$ (denoted as $T_j \Rightarrow T_i$) to make the schedule serializable, as otherwise, the read of $T_j$ would be outdated. Unfortunately, the problem is typically more complex as \textit{cycles of conflicts} can occur, such that simple reordering can not resolve the problem. For example, if we have the cycle of conflicts $T_i \nrightarrow T_j \nrightarrow T_k \nrightarrow T_i$, there is no order of these three transactions that is serializable. 
%In this case, the set of transactions is said to be \textbf{not} \textit{conflict serializable}. 
Therefore, before reordering transactions, our mechanism must actually first remove certain transactions of~$S$ to form a subset~$S' \subseteq S$, from which a serializable schedule can be generated. Of course, a goal must be to remove as few transactions as possible. Finally, after computing $S'$, we can derive a concrete serializable schedule from the transactions in $S'$. 

On a high-level, we have to carry out the steps as shown in the pseudo-code of Algorithm~\ref{alg:ordering} to create a serializable schedule for a set of transactions~$S$.

\begin{figure}[htb!]
\lstset{style=MYC}
\begin{lstlisting}[escapeinside={(*}{*)}]
func reordering(Transaction[] S) {
   // Step 1: Inspect the read/write set of all transactions and build a conflict graph.
   (*\label{l:buildConflictGraph}*)Graph cg = buildConflictGraph(S) 
   // Step 2: Within cg, we have to identify all occurring cycles. Divide cg into 
   // strongly connected subgraphs using Tarjans algorithm [2] in divideIntoSubgraphs(). 
   (*\label{l:divideIntoSubgraphs}*)Graph[] cg_subgraphs = divideIntoSubgraphs(cg)
   // Each strongly connected subgraph of cg with more than one node must contain at 
   // least one cycle. We identify the cycles within the subgraphs using Johnsons 
   // algorithm [1] in getAllCycles().
   (*\label{l:step3start}*)Cycle[] cycles = emptyList()
   foreach subgraph in cg_subgraphs:
      if(subgraph.numNodes() > 1):
         (*\label{l:step3end}*)cycles.add(subgraph.getAllCycles())
   // Step 3: To remove the cycles in cg, we have to remove conflicting transactions from 
   // S. For each transaction of S, we count in how many cycles it occurs.
   MaxHeap transactions_in_cycles = emptyMaxHeap()
   foreach Cycle c in cycles:
      foreach Transaction t in c:
         if transactions_in_cycles.contains(t)
            transactions_in_cycles[t]++
         else
            transactions_in_cycles[t] = 1
   // Step 4: Let us define S(*'*) as S. We now greedily remove the transaction from S(*'*) that 
   // occurs in most cycles, until all cycles have been resolved.
   (*\label{l:step4start}*)Transaction[] S(*'*) = S
   while not cycles.empty():
      Transaction t = transactions_in_cycles.popMax()
      S(*'*).remove(t)
      foreach Cycle c in cycles:
         if c.contains(t):
            c.remove(t)
            (*\label{l:step4end}*)cycles.remove(c)
         foreach Transaction t(*'*) in c:
            transactions_in_cycles[t(*'*)]--
   // Step 5: From S(*'*) we have to form the actual serializable schedule. We start by 
   // building the (cycle-free) conflict graph of S(*'*). 
   (*\label{l:buildConflictGraph2}*)Graph cg(*'*) = buildConflictGraph(S(*'*))
   // Compute schedule. We start at some node of the graph, that hasn't been visited yet.
   Transactions[] order = emptyList() 
   Node startNode = cg(*'*).getNextNode()
   while order.length() < cg(*'*).numNodes():
      addNode = true
      (*\label{l:searchSourceStart}*)if startNode.alreadyScheduled():
         startNode = cg(*'*).getNextNode()
         continue
      // Traverse upwards to find a source
      foreach Node parentNode in startNode.parents():
         if not parentNode.alreadyScheduled(): 
            startNode = parentNode
            addNode = false
            (*\label{l:searchSourceEnd}*)break
      // A source has been found, so schedule it and traverse downwards.
      (*\label{l:scheduleStart}*)if addNode:
         startNode.scheduled()
         order.append(startNode)
         foreach Node childNode in startNode.children():
            if not childNode.alreadyScheduled():
               startNode = childNode
               (*\label{l:scheduleEnd}*)break
   // we invert the order to get the actual schedule
   return order.invert()
}
\end{lstlisting}
\vspace{-1.2cm}
\caption{Pseudo code of the reordering mechanism.}
\label{alg:ordering}
\end{figure}

\begin{table}[!htb]
%\vspace{-0.3cm}
	%\scriptsize
	\setlength{\tabcolsep}{3pt}
	\begin{center}
		\caption{Ten unique keys that are accessed by six transactions, separated in read set and write set.}
		%\vspace{-0.4cm}
		\begin{tabular}{ C{4cm} || C{0.6cm}  C{0.6cm}  C{0.6cm}  C{0.6cm} C{0.6cm}  C{0.6cm}  C{0.6cm}  C{0.6cm}  C{0.6cm}  C{0.6cm}}
			\hline
			 &  \multicolumn{10}{c}{\textbf{Read Set}} \\
			\textbf{Transactions} &  $K_0$ & $K_1$ & $K_2$ & $K_3 $ & $K_4$ & $K_5$ & $K_6$ & $K_7$ & $K_8$ & $K_9$ \\\hline\hline
			$T_0$ & \textbf{1} & \textbf{1} & 0 & 0 & 0 & 0 & 0 & 0 & 0 & 0\\
			$T_1$ & 0 & 0 & 0 & \textbf{1} & \textbf{1} & \textbf{1} & 0 & 0 & 0 & 0\\
			$T_2$ & 0 & 0 & 0 & 0 & 0 & 0 & \textbf{1} & \textbf{1} & 0 & 0\\
			$T_3$ & 0 & 0 & \textbf{1} & 0 & 0 & 0 & 0 & 0 & \textbf{1} & 0\\
			$T_4$ & 0 & 0 & 0 & 0 & 0 & 0 & 0 & 0 & 0 & \textbf{1}\\
			$T_5$ & 0 & 0 & 0 & 0 & 0 & 0 & 0 & 0 & 0 & 0\\
			\hline
			\multicolumn{11}{c}{}\\
			\hline
			 &  \multicolumn{10}{c}{\textbf{Write Set}}\\
			
			 \textbf{Transactions} &  $K_0$ & $K_1$ & $K_2$ & $K_3 $ & $K_4$ & $K_5$ & $K_6$ & $K_7$ & $K_8$ & $K_9$ \\\hline\hline
			$T_0$ & 0 & 0 & \textbf{1} & 0 & 0 & 0 & 0 & 0 & 0 & 0\\
			$T_1$ & \textbf{1} & 0 & 0 & 0 & 0 & 0 & 0 & 0 & 0 & 0\\
			$T_2$ & 0 & 0 & 0 & \textbf{1} & 0 & 0 & 0 & 0 & 0 & \textbf{1}\\
			$T_3$ & 0 & \textbf{1} & 0 & 0 & \textbf{1} & 0 & 0 & 0 & 0 & 0\\
			$T_4$ & 0 & 0 & 0 & 0 & 0 & \textbf{1} & \textbf{1} & 0 & \textbf{1} & 0\\
			$T_5$ & 0 & 0 & 0 & 0 & 0 & 0 & 0 & \textbf{1} & 0 & 0\\
			\hline
		\end{tabular}
		\label{table:conflictkeys}
	\end{center}
\end{table}

\subsubsection{Example}

To understand the principle and to discuss some of the implementation details, let us go through a concrete example. Let us assume we have a set~$S$ of six transactions~$T_0$ to~$T_5$ to consider for reordering. These six transactions have read and write sets as shown in Table~\ref{table:conflictkeys}. In total, they access ten unique keys~$K_0$ to~$K_9$. 

\begin{figure}[!htb]
    \begin{center}
        \includegraphics[width=6cm]{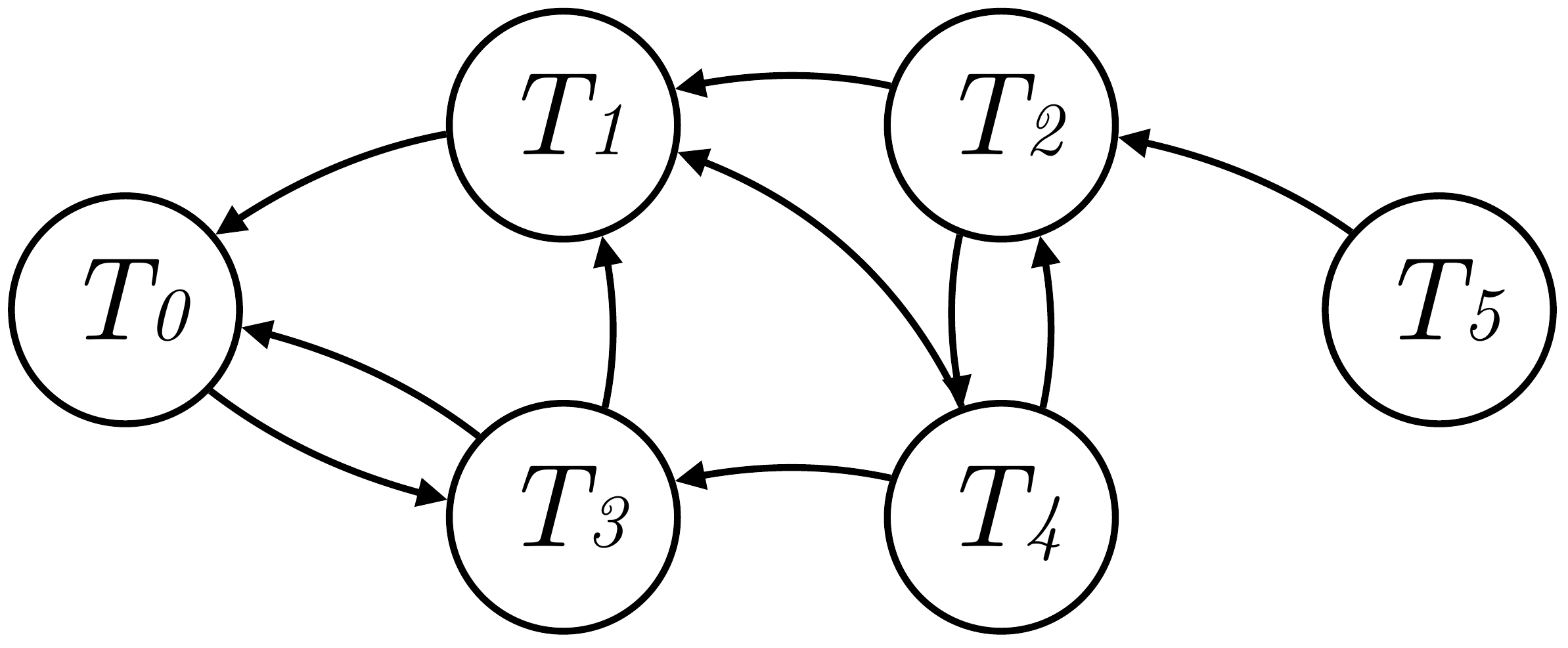}
    \end{center}
    \caption{Conflict graph $C(S)$ of the transactions in~$S$.}
    \label{figs:fabric++_ordering}
\end{figure}

\textbf{Step (1)}: Based on this information, we now have to generate the conflict graph of the transactions as done by the function \smalltt{buildConflictGraph()} in line~\ref{l:buildConflictGraph} of Algorithm~\ref{alg:ordering}. To do so in an efficient way, we interpret the rows of Table~\ref{table:conflictkeys} as bit-vectors of length~$10$. Let us refer to them as $vec_r(T_i)$ for the reading accesses and $vec_w(T_i)$ for the writing accesses of a transaction~$T_i$. For each transaction~$T_i$, we now perform a bitwise \texttt{\&}-operation between $vec_r(T_i)$ and $vec_w(T_j)$ for all $j \neq i$. If the result of an \texttt{\&}-operation is not~$0$, we have identified a read-write conflict and create an edge in the conflict graph between the corresponding transactions. For example, for $T_0$ we have the reading accesses $vec_r(T_0) = 1100000000$
 The bitwise \texttt{\&}-operation with $vec_w(T_1) = 1000000000$ leads to $ vec_r(T_0) \texttt{ \& }  vec_w(T_1) = 1000000000$, which is not~$0$. This means $T_1$ writes a key that $T_0$ is reading and thus, we put a corresponding edge in the conflict graph.  
As a result, we obtain the conflict graph~$C(S)$ of our six transactions as shown in Figure~\ref{figs:fabric++_ordering}.

\noindent Note that this algorithm has quadratic complexity on the number of transactions. Still, we apply it as the number of transactions to consider is very small in practice due to the limitation by the block size and therefore, the overhead is negligible.

\begin{figure}[!htb]
    \begin{center}
        \includegraphics[width=6cm]{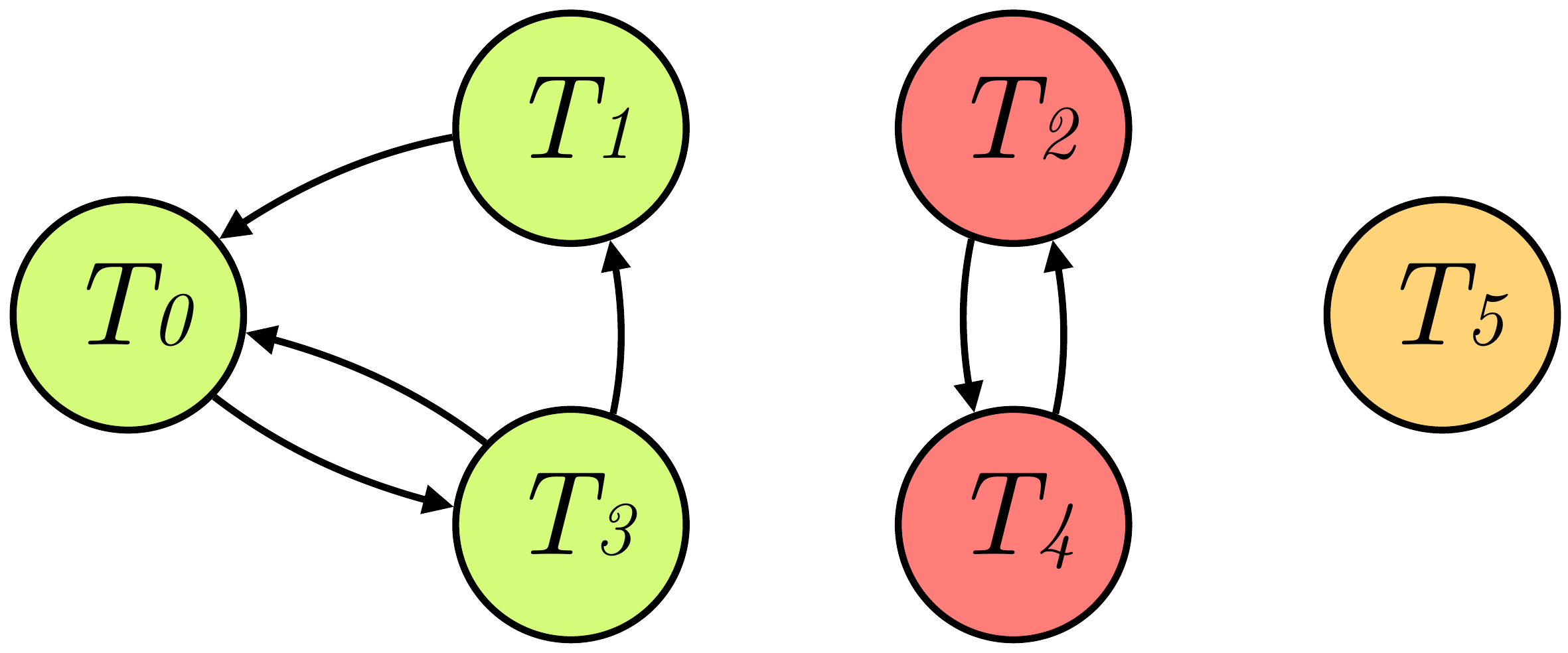}
    \end{center}
    \caption{The three strongly connected subgraphs of the conflict graph of Figure~\ref{figs:fabric++_ordering}.}
    \label{figs:fabric++_ordering_subgraphs}
\end{figure}

\textbf{Step~(2)}: To identify the cycles, we apply Tarjan's algorithm~\cite{Tarjan72} in the function~\smalltt{divideIntoSubgraphs()} in line~\ref{l:divideIntoSubgraphs} to identify all strongly connected subgraphs. In general, this can be done in linear time in $\mathcal{O}(N+E)$ over the number of nodes~$N$ and number of edges~$E$ and results in the three subgraphs as shown in Figure~\ref{figs:fabric++_ordering_subgraphs}. 

Using Johnson's algorithm~\cite{Johnson75}, we then identify all cycles within the strongly connected subgraphs. Again, this step can be done in linear time in $\mathcal{O}((N+E) \cdot (C+1))$, where $C$ is the number of cycles. Therefore, if there are no cycles in the subgraphs, the overhead of this step is very small. 

We identify that the first subgraph (colored in green) consisting of~$T_0$, $T_1$, and $T_3$ contains the two cycles~\mbox{$c_1=T_0 \nrightarrow T_3 \nrightarrow T_0$} and~$c_2=T_0 \nrightarrow T_3 \nrightarrow T_1 \nrightarrow T_0$. The second subgraph (colored in red) consisting of~$T_2$ and $T_4$ contains the cycle~$c_3=T_2 \nrightarrow T_4 \nrightarrow T_2$. The third subgraph (colored in yellow) contains only one node and is thus cycle-free.

\textbf{Step~(3)}: From this information, we can build a table denoting for every transaction in which cycle it appears as shown in the lines~\ref{l:step3start}~to~\ref{l:step3end} of Algorithm~\ref{alg:ordering}. Table~\ref{table:fabric++_ordering_cycles} visualizes the result for our example. If a transaction~$T_i$ is part of a cycle~$c_j$, the corresponding cell is set to~$1$, otherwise~$0$. The last row of the table sums up for every transaction in how many cycles it is contained in total.

\begin{table}[!htb]
%\vspace{-0.3cm}
	%\scriptsize
	\setlength{\tabcolsep}{3pt}
	\begin{center}
		\caption{If a transaction~$T_i$ is a part of a cycle~$c_j$, the corresponding cell is set to~1, otherwise~0. The last row contains for every transaction the total number of cycles, in which it appears.}
		%\vspace{-0.4cm}
		\begin{tabular}{ C{1.5cm} || C{0.9cm} | C{0.9cm} | C{0.9cm} | C{0.9cm} | C{0.9cm} | C{0.9cm} }
			\hline
			\textbf{Cycle} & $T_0$ & $T_1$ & $T_2$ & $T_3 $ & $T_4$ & $T_5$\\
			\hline
			$c_1$ & 1 & 0 & 0 & 1 & 0 & 0\\\hline
			$c_2$ & 1 & 1 & 0 & 1 & 0 & 0\\\hline
			$c_3$ & 0 & 0 & 1 & 0 & 1 & 0\\\hline\hline
			$\sum$ & 2 & 1 & 1 & 2 & 1 & 0\\
			\hline
		\end{tabular}
		\label{table:fabric++_ordering_cycles}
	\end{center}
\end{table}

\textbf{Step~(4)}: We now iteratively remove transactions, that participate in cycles, starting from the ones that appear in most cycles. The lines~\ref{l:step4start}~to~\ref{l:step4end} of Algorithm~\ref{alg:ordering} show the corresponding pseudo-code. As we can see, $T_0$ and $T_3$ both appear in two cycles, so we take care of them first. If we can choose between two transactions, such as $T_0$ and $T_3$, we pick the one with the smaller subscript. This assures that our algorithm is deterministic. We remove~$T_0$, which clears all cycles in which $T_0$ appears, namely~$c_1$ and $c_2$. The sum is updated accordingly, as we can see in Table~\ref{table:fabric++_ordering_cycles2}.

\begin{table}[!htb]
%\vspace{-0.3cm}
	%\scriptsize
	\setlength{\tabcolsep}{3pt}
	\begin{center}
		\caption{Removing $T_0$ clears the cycles~$c_1$ and $c_2$.}
		%\vspace{-0.4cm}
		\begin{tabular}{ C{1.5cm} || C{0.9cm} | C{0.9cm} | C{0.9cm} | C{0.9cm} | C{0.9cm} | C{0.9cm} }
			\hline
			\textbf{Cycle} & \cancel{$T_0$ } & $T_1$ & $T_2$ & $T_3 $ & $T_4$ & $T_5$\\
			\hline
			\cancel{$c_1$} & \cancel{ 1 } & \cancel{ 0 } & \cancel{ 0 } & \cancel{ 1 } & \cancel{ 0 } & \cancel{ 0 }\\\hline
			\cancel{$c_2$} & \cancel{ 1 } & \cancel{ 1 } & \cancel{ 0 } & \cancel{ 1 } & \cancel{ 0 } & \cancel{ 0 }\\\hline
			$c_3$ & 0 & 0 & 1 & 0 & 1 & 0\\\hline\hline
			$\sum$ & 0 & 0 & 1 & 0 & 1 & 0\\
			\hline
		\end{tabular}
		\label{table:fabric++_ordering_cycles2}
	\end{center}
	%\vspace{-0.4cm}
\end{table}

\noindent The transactions~$T_2$ and $T_4$ remain with a participation in cycle~$c_3$ each. We remove~$T_2$ which clears~$c_3$ and thereby the last cycle. This results in the state of Table~\ref{table:fabric++_ordering_cycles3}.

\begin{table}[!htb]
%\vspace{-0.3cm}
	%\scriptsize
	\setlength{\tabcolsep}{3pt}
	\begin{center}
		\caption{Removing $T_2$ clears the last cycle~$c_3$.}
		%\vspace{-0.4cm}
		\begin{tabular}{ C{1.5cm} || C{0.9cm} | C{0.9cm} | C{0.9cm} | C{0.9cm} | C{0.9cm} | C{0.9cm} }
			\hline
			\textbf{Cycle} & \cancel{$T_0$ } & $T_1$ & \cancel{$T_2$ } & $T_3 $ & $T_4$ & $T_5$\\
			\hline
			\cancel{$c_1$} & \cancel{ 1 } & \cancel{ 0 } & \cancel{ 0 } & \cancel{ 1 } & \cancel{ 0 } & \cancel{ 0 }\\\hline
			\cancel{$c_2$} & \cancel{ 1 } & \cancel{ 1 } & \cancel{ 0 } & \cancel{ 1 } & \cancel{ 0 } & \cancel{ 0 }\\\hline
			\cancel{$c_3$} & \cancel{ 0 } & \cancel{ 0 } & \cancel{ 1 } & \cancel{ 0 } & \cancel{ 1 } & \cancel{ 0 }\\\hline\hline
			$\sum$ & 0 & 0 & 0 & 0 & 0 & 0\\
			\hline
		\end{tabular}
		\label{table:fabric++_ordering_cycles3}
	\end{center}
	%\vspace{-0.4cm}
\end{table}

\begin{figure}[!htb]
    \begin{center}
        \includegraphics[width=6cm]{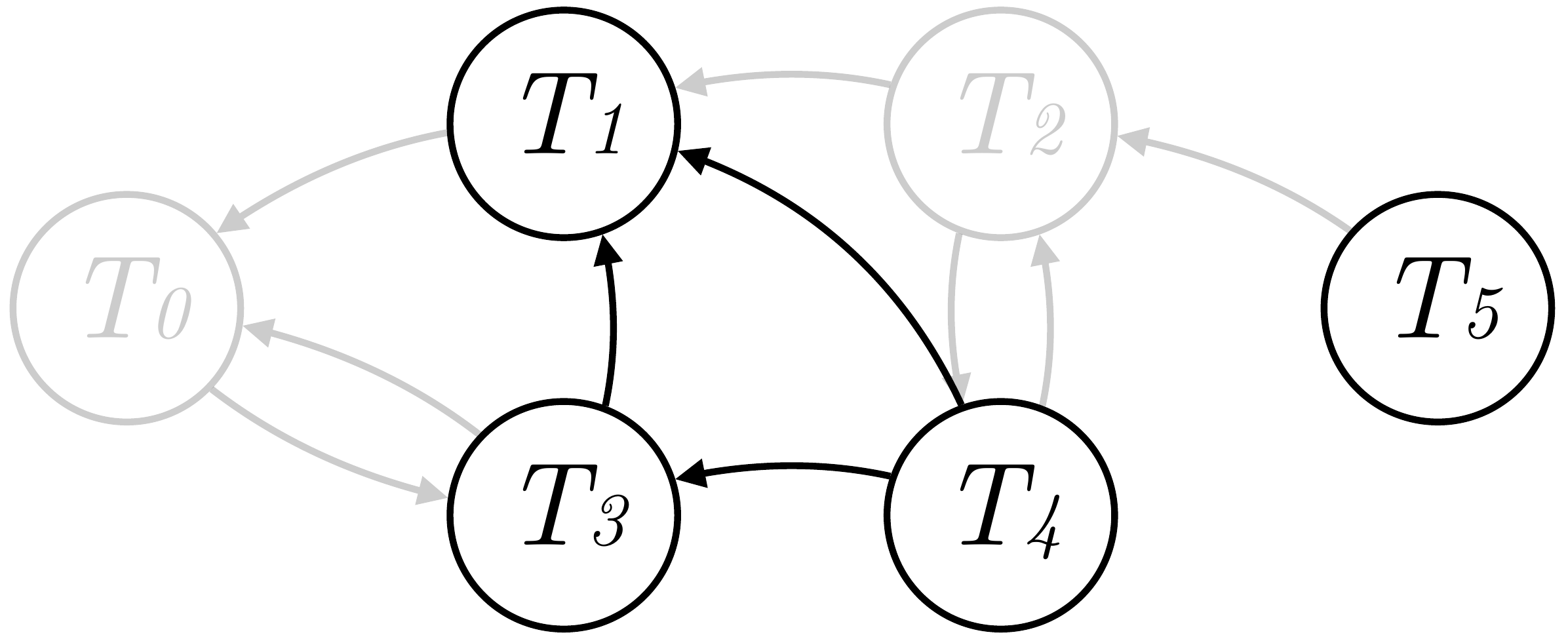}
    \end{center}
    \caption{The cycle-free conflict graph~$C(S')$, containing only the transactions $T_1$, $T_3$, $T_4$, and $T_5$.}
    \label{figs:fabric++_ordering_cyclefree}
\end{figure}

\noindent From this we now know that from the set~$S'=\{T_1,T_3,T_4,T_5\}$ we can generate a serializable schedule, leading to the cycle-free conflict graph~$C(S')$ (line~\ref{l:buildConflictGraph2}) as shown in Figure~\ref{figs:fabric++_ordering_cyclefree}. 

\textbf{Step~(5)}: Generating the final schedule is essentially a repetitive execution of two parts until all nodes are scheduled: (a) the locating of the source node in the current subgraph (lines~\ref{l:searchSourceStart}~to~\ref{l:searchSourceEnd}) and (b) the scheduling of all nodes that reachable from that source (lines~\ref{l:scheduleStart}~to~\ref{l:scheduleEnd}).

 We start part~(a) at the node of $C(S')$ representing the transaction with the smallest subscript, namely~$T_1$. From this starting node, we have to find a source node, as sources have to be scheduled last. $T_1$ has two parents, namely~$T_3$ and $T_4$, so it not a source. We follow the edge to $T_3$, which has not been visited yet but is also not a source, as it has~$T_4$ as a parent as well. We follow the edge to~$T_4$, which has not been visited yet and which is a source. Therefore, we can schedule~$T_4$ safely at the last position in our schedule, to which we refer to as  \textit{position~$4$}. Now, part~(b) starts as all nodes that are reachable from~$T_4$ must be scheduled before it. $T_4$ has two children, namely~$T_1$ and $T_3$. We follow the edge to~$T_1$, which has not been scheduled yet. However, as~$T_1$ has an incoming edge from~$T_3$, we also can not directly schedule it. First, we visit~$T_3$ and identify that it has a parent in form of~$T_4$, the source at which we started. With this information, we know that $T_3$ must be scheduled at position~$3$ and $T_1$ must be scheduled at position~$2$. This ends part~(b), as all reachable nodes have been scheduled. 
 Next, we restart at the only remaining node~$T_5$. As $T_5$ is not only a source but also a sink, we can schedule it instantly at position~$1$. This results in the final schedule~$T_5 \Rightarrow T_1 \Rightarrow T_3 \Rightarrow T_4$, which is returned to the orderer.
 
 Please note that our reordering mechanism is not guaranteed to abort a minimal number of transactions, as this would be a NP-hard problem. However, it offers a very lightweight way to generate a serializable schedule with a small number of aborts.

\subsubsection{Batch Cutting}

In the context of transaction reordering, we have to discuss and extend a mechanism within the ordering service, that we omitted for simplicity in the description of Fabric in Section~\ref{sec:fabric}, namely \textit{batch cutting}. When the ordering service receives the transactions in form of a constant stream, it decides based on multiple criteria when to "cut" a batch of transactions to finalize it and to form the block. In the vanilla version, a batch is cut as soon as one of the following three conditions hold: (a) The batch contains a certain number of transactions. (b) The batch has reached a certain size in terms of bytes. (c) A certain amount of time has passed since the first transaction of this batch was received. 

In Fabric++, we extend these criteria by one additional condition. We also cut the batch, if (d) the transactions within the batch access a certain number of unique keys. This condition ensures that the runtime of our reordering mechanism, in particular the time of step~(1), remains bounded.

\subsubsection{Micro-Benchmark}

\begin{figure}[!htb]
    \begin{center}
        \includegraphics[width=10cm]{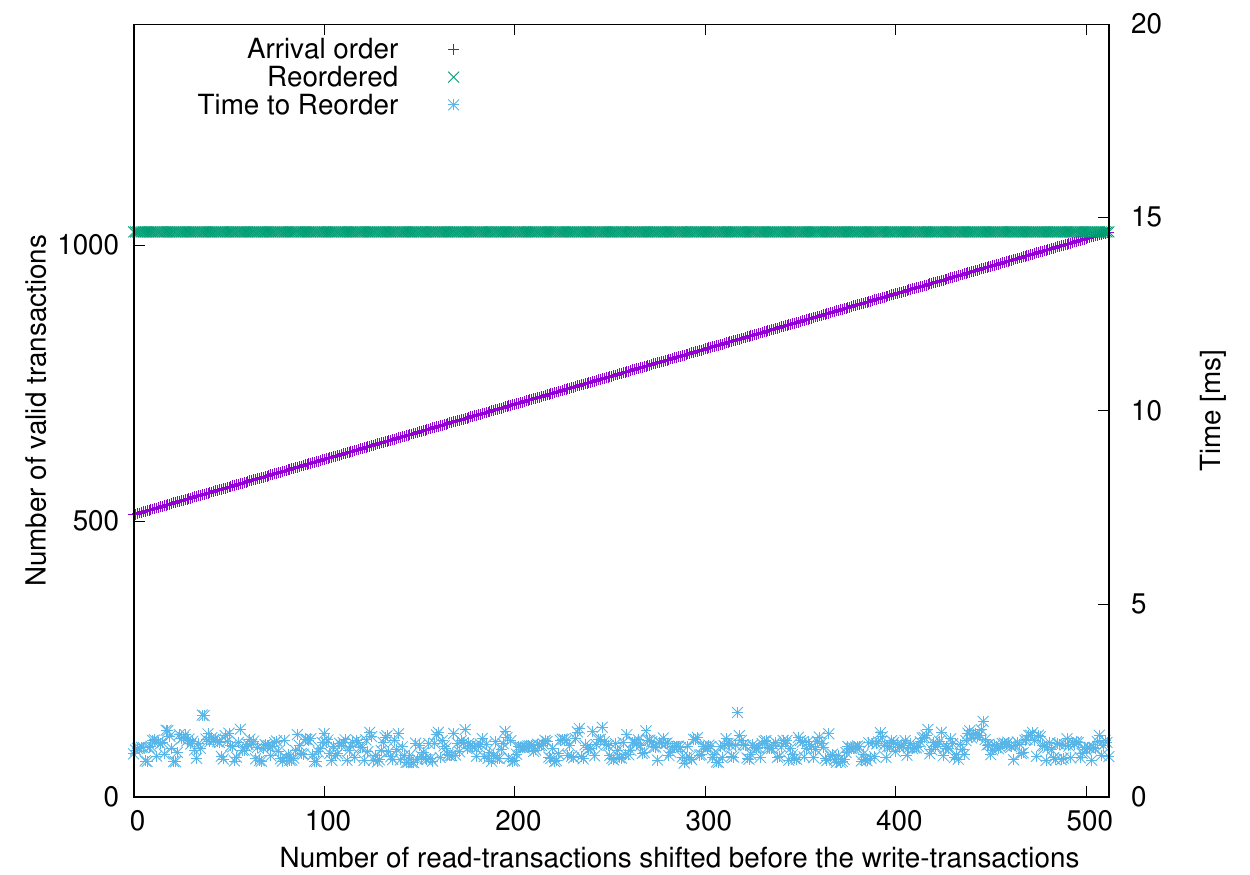}
    \end{center}
    \caption{Workload 1: Varying the number of conflicts.}
    \label{figs:ordering_benchmark1}
\end{figure}

To analyze the effectiveness of our reordering mechanism, we first evaluate it in a stand-alone micro-benchmark in isolation of Fabric. For a given sequence of input transactions we compute the number of valid transactions for this particular sequence (called "arrival order" in the following plots) as well as for the sequence that is generated by our reordering mechanism (called "reordered" in the following plots). Additionally, we measure the time to compute the reordered schedule. In Figure~\ref{figs:ordering_benchmark1}, we test a workload pattern with varying number of conflicts. For the interested reader, we provide a second micro-benchmark in the Appendix~\ref{sec:ordering_benchmark2} on the effect of varying the length of the cycles (Figure~\ref{figs:ordering_benchmark2}) and see how well our reordering mechanism performs in comparison to the naive arrival order.

\subsubsection{Micro-Benchmark 1: Interleave reads and writes to vary the number of conflicts} 

The first input sequence we test consists of two equal sized sub sequences, where one subsequence contains only  transactions that perform writes (colored in \textcolor{red}{red}) and the other sequence only transactions that read (colored in \textcolor{blue}{blue}). Each transaction performs only one operation (either read or write). Neither two writes nor two reads happen to the same key. For the example of $n=6$~transactions, we start with the following sequence~$S_1$:
\begin{small}
$$ S_1 = T[\textcolor{red}{w(k_1)}], T[\textcolor{red}{w(k_2)}],T[\textcolor{red}{w(k_3)}], T[\textcolor{blue}{r(k_1)}], T[\textcolor{blue}{r(k_2)}], T[\textcolor{blue}{r(k_3)}] $$
\end{small}To generate $S_i$, we move the last transaction of~$S_{i-1}$ to the front, leading to the following sequences~$S_2$, $S_3$, and $S_4$.
\begin{small}
$$ S_2 = T[\textcolor{blue}{r(k_3)}], T[\textcolor{red}{w(k_1)}], T[\textcolor{red}{w(k_2)}],T[\textcolor{red}{w(k_3)}], T[\textcolor{blue}{r(k_1)}], T[\textcolor{blue}{r(k_2)}]$$
$$ S_3 = T[\textcolor{blue}{r(k_2)}], T[\textcolor{blue}{r(k_3)}], T[\textcolor{red}{w(k_1)}], T[\textcolor{red}{w(k_2)}],T[\textcolor{red}{w(k_3)}], T[\textcolor{blue}{r(k_1)}]$$
$$ S_4 = T[\textcolor{blue}{r(k_1)}], T[\textcolor{blue}{r(k_2)}], T[\textcolor{blue}{r(k_3)}], T[\textcolor{red}{w(k_1)}], T[\textcolor{red}{w(k_2)}],T[\textcolor{red}{w(k_3)}]$$
\end{small}The more writing transactions happen before the corresponding reading transactions, the more conflicts happen. We want to find out whether our reordering mechanism can solve this problem. 

Figure~\ref{figs:ordering_benchmark1} shows the results for~$n=1024$~transactions.
As we can see, our reordering mechanism is able to reorder the transactions for every input sequence in a way such that all transactions are valid. In contrast to that, the arrival order suffers under a lot of invalid reading transactions, if writing transactions happen before. We can also see that our reordering mechanism is computationally cheap: it takes only around 1 to 2~ms to rearrange the transactions on a Macbook Pro with Intel Core i7 running at 3.1~GHz.

\subsection{Early Transaction Abort using Advanced Concurrency Control}
\label{ssec:early_abort}

% Looking back: early abort in ordering
The reordering mechanism previously described not only tries to minimize the number of unnecessary aborts, it also enables a form of \textit{early abort}. Transactions, that are removed from~$S$ because of their participation in conflict cycles can be aborted already in the ordering phase instead of later on the validation phase. This assures that less transactions are distributed across the network. 

% Pushing it further
In the following, we want to push this concept of aborting transactions as early as possible in the pipeline to the limits. Additionally to early aborting transactions that occur in conflict cycles, we can integrate two more applications of early abort, as we will describe in Section~\ref{ssec:early_abort_simulation} and Section~\ref{ssec:early_abort_ordering}. The first one is happening already in the simulation phase. Let us see in the following how this works.

\subsubsection{Early Abort in the Simulation Phase} 
\label{ssec:early_abort_simulation}

To realize early abort in the simulation phase, we first have to extend Fabric by a more fine-grained concurrency control mechanism, that allows for the parallel execution of simulation and validation phase within a peer. With such a mechanism at hand, we have the chance of identifying stale reads \textit{during} the simulation already. 

To understand the concept, let us consider the example from Section~\ref{sssec:early_abort_simulation} again. With a fine-grained concurrency control mechanism, the block containing~$T_1$, $T_2$, $T_3$, and $T_4$ would not have to pend for validation while the smart contract bound to the proposal~$T_5$ is simulating. Instead, the four transactions would apply their updates in an atomic fashion \textit{while} $T_5$ is simulating. As a consequence of this design, for every read $T_5$ performs, we can check whether the read value is still up-to-date. As soon as we detect a stale read, we can abort the simulation of the transaction proposal. Additionally, we directly notify the corresponding client about the abort, such that it can resubmit the proposal without delay.

Let us discuss in the following, how exactly our fine-grained concurrency control mechanism works and how we realize it in Fabric++.
In the context of modern database systems, advanced concurrency control mechanisms are well established~\cite{hyper, fsfekete, movcc, gconemvcc, gctwomvcc, anker}. Instead of locking the entire store, these techniques typically perform a fine-grained locking on the record level or even at the level of individual cells/values. As there is conceptually no difference between the store of a database system and the store used within the Fabric peers,  similar techniques can be applied here. 

\begin{figure}[!htb]
    \begin{center}
        \includegraphics[width=14cm]{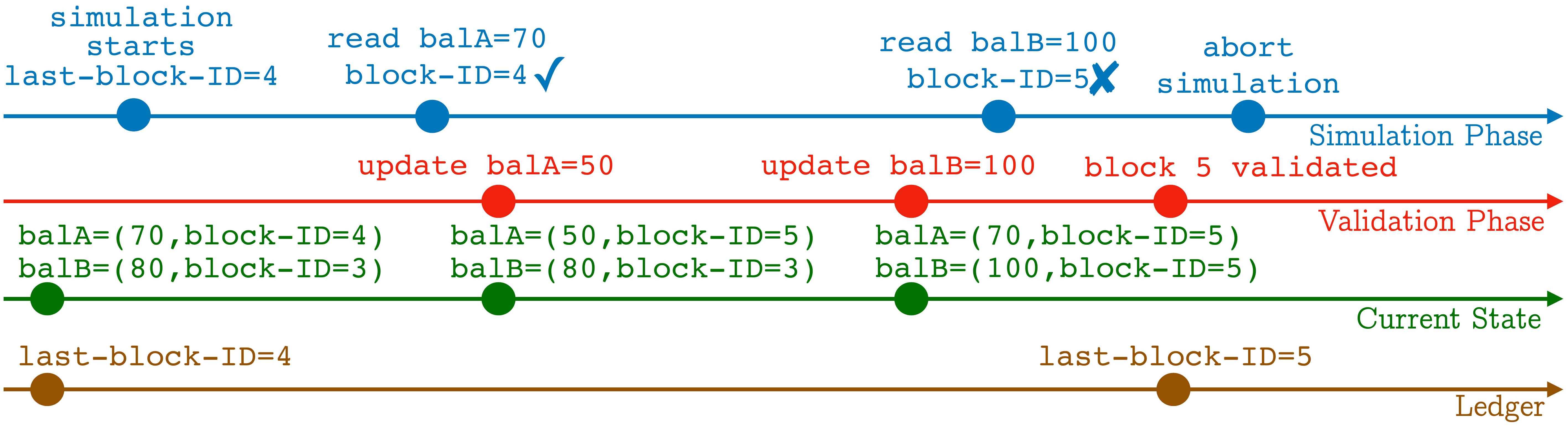}
    \end{center}
    \caption{Parallelization with early abort using our fine-grained concurrency control.}
    \label{figs:fabric++_cc}
\end{figure} 

As discussed in Section~\ref{sec:fabric}, Fabric implements its current state in form of a key-value store, which maps each individual key to a pair of value and version-number. The version-number is actually composed of the ID of the transaction, that performed the update, as well as the ID of the block that contains the transaction. 
In the original version of Fabric, the sole purpose of the version-numbers is to identify stale reads. In the validation phase, for every transaction we check whether the version-number of the read value still matches the one in the current state. 

We can go one step further and exploit the available version-numbers to implement a lock-free concurrency control mechanism protecting the current state. 
To do so, in Fabric++, we first remove the read-write lock, that was unnecessarily sequentializing simulation and validation phase. The version-number, that is maintained with each value, is sufficient to ensure the same transaction isolation semantics as the vanilla version. 
As no lock is acquired anymore, we need a mechanism to ensure that updates performed by the validation phase are not seen by simulation phases running in parallel.
To achieve this behavior, during simulation, we have to inspect the version-number of every read value and test whether it is still up-to-date. 

Figure~\ref{figs:fabric++_cc} visualizes this concept using a concrete example. 
At the start of the simulation phase, we first identify the \smalltt{block-ID} of the last block that made it into the ledger. Let us refer to this \smalltt{block-ID} as the \smalltt{last-block-ID}. In our example, \smalltt{last-block-ID~=~4}. During the simulation of a smart contract bound to a transaction proposal~$T_{exec}$, no read must encounter a version-number containing a \smalltt{block-ID} higher than the \smalltt{last-block-ID}. If it does see a higher \smalltt{block-ID} it means that during the simulation phase, a validated transaction~$T_{valid}$ in the validation phase modified a value in the read set of~$T_{exec}$ and thus, the read set is outdated.

In our example, the read of $balA=70$ in the simulation phase happens \textit{before} the update of $balA$ to~$50$ in the validation phase. This is reflected by the version-number of $balA$, namely~\smalltt{block-ID~=~4}. Therefore, this read is up-to-date and the simulation continues. In contrast to that, the read of $balB$ happens \textit{after} the update of $balB$~to~$100$ in the validation phase. This is reflected by the version-number of $balB$, namely~\smalltt{block-ID~=~5}. As \smalltt{5}~is higher than the \smalltt{last-block-ID~=~4}, we can directly classify~$T_{exec}$ as invalid, as the transaction will not have a chance to pass the validation phase later on. Please note that the overall correctness of our lock-free mechanism is ensured by the atomic updates of the version-numbers. 

\subsubsection{Early Abort in the Ordering Phase}
\label{ssec:early_abort_ordering}
In addition to the early abort in the simulation phase, as explained in Section~\ref{ssec:early_abort_simulation}, we can transition a similar concept also to the ordering phase. As Fabric performs commits at the granularity of whole blocks, two transactions within the same block, that read the same key, must read the same version of that key. For example, let us consider two transactions~$T_6$ and $T_7$, where $T_6$ is ordered before $T_7$ within the same block ($T_6 \Rightarrow T_7$). If $T_6$ read version~$v_1$ of a key~$k$ and $T_7$ read version~$v_2$ of $k$ in their respective simulations, then $T_7$ is invalid. Such a version mismatch can happen, if between the simulations of $T_6$ and $T_7$ a change to the value of~$k$ was committed by a valid transaction from a previous block.  
Therefore, as soon as we detect a version mismatch between transactions within the same block, we can early abort the latter transaction. Again, this strategy assures that only those transactions end up in a block, that have a realistic chance of commit.  

\section{Experimental Evaluation}
\label{sec:ea}

In the previous section, we have extended and modified core components of Fabric in several ways, turning it into Fabric++. It is now time to evaluate the modifications in terms of effectiveness. Primarily, we are interested in the throughput of valid/successful and invalid/failed transactions, that make it through the system. Secondarily, we are interested on the influence of certain system configurations and the workload characteristics on the system.

\subsection{Setup}

Before starting with the actual experiments, let us discuss the setup. Our cluster consists of six identical servers, that are located within the same rack and connected via gigabit-ethernet.  Four machines serve as peers, one machine runs the ordering service, and one machine serves as the client, which fires transaction proposals. 
Each server consists of two quad-core Intel Xeon CPU E5-2407 (SandyBridge architecture) running at $2.2$~GHz with $32$KB of L1 cache, $256$KB of L2 cache, and $10$MB of a shared L3 cache. $24$GB of DDR3 ram are attached to each of the two NUMA regions. The operating system used is a 64-bit Arch Linux with kernel version~$4.17$. Fabric is set up to use LevelDB as the current state database.

\subsection{Benchmark Framework and Workload}

In the database community, there exist numerous established benchmarks that can be used to test and to compare systems, such as TPC-C~\cite{tpcc}, TPC-H~\cite{tpch}, or YCSB~\cite{ycsb}. Unfortunately, since blockchains are still a relatively young field, there exist only very few benchmarks with standardized workloads. 

At first, we looked at the Caliper~\cite{caliper} benchmarking suite which seemed like a natural candidate, as it is part of the Hyperledger project just like Fabric. It is compatible to Fabric~$1.2$, but comes with a few limitations: First, the framework provides only sample smart contracts and not a real benchmarking workload.  Second, for certain metrics such as transactions per second or latency, it remains unspecified how they are actually measured. Third, it supports only a single channel. Apart from these limitations, other researcher have experienced incorrect behavior of Caliper in form of events, that were not properly registered.~\cite{gauge_changes}. As a consequence, they released a fork of Caliper named Gauge~\cite{gauge} that claims to resolve these problems. Unfortunately, Gauge is not compatible with version~$1.2$ of Fabric right now. 
Next, we looked at Blockbench, which originates from a survey paper~\cite{BlockchainSurvey} on blockchain systems. While Blockbench actually provides some benchmarking workloads such as YCSB, again, it lacks the support for Fabric~$1.2$ and would need significant changes to make it compatible.

As a consequence of this journey, we decided to build our own benchmarking framework and to introduce a highly customizable workload. This allows us to fire transaction proposals at a specified rate from multiple clients in multiple channels. 
Our benchmark setup looks as follows: Initially, we create a certain number of accounts (10,000 accounts throughout this section, 20,000 accounts in Appendix~\ref{sec:extended_throughput}), each represented by a randomly generated account balance. Our workload is formed of a single smart contract, that reads and writes an adjustable number of account balances, simulating a typical asset transfer scenario between accounts. Among the accounts, there exist a certain number of hot accounts, that are involved in transactions more frequently than the remaining ones. By varying the number of read and write accesses per transaction, the probability of picking hot accounts, and the number of hot accounts, we are able to generate a wide range of different workloads.  

In a single run, we fire a constant stream of transaction proposals, that are bound to our smart contract, for a certain amount of time at a certain firing rate. In the following, we test numerous different system and workload configurations to identify the impact of the system. In the individual experiments, we will detail the chosen configuration.

\subsection{Transactional Throughput}
\label{ssec:throughput}
\vspace{-0.2cm}

\begin{figure*}[!htb]
    \vspace{-0.2cm}
    \begin{center}
        \includegraphics[width=\textwidth, height=21cm, trim={2.8cm 0 1.2cm 0}, clip]{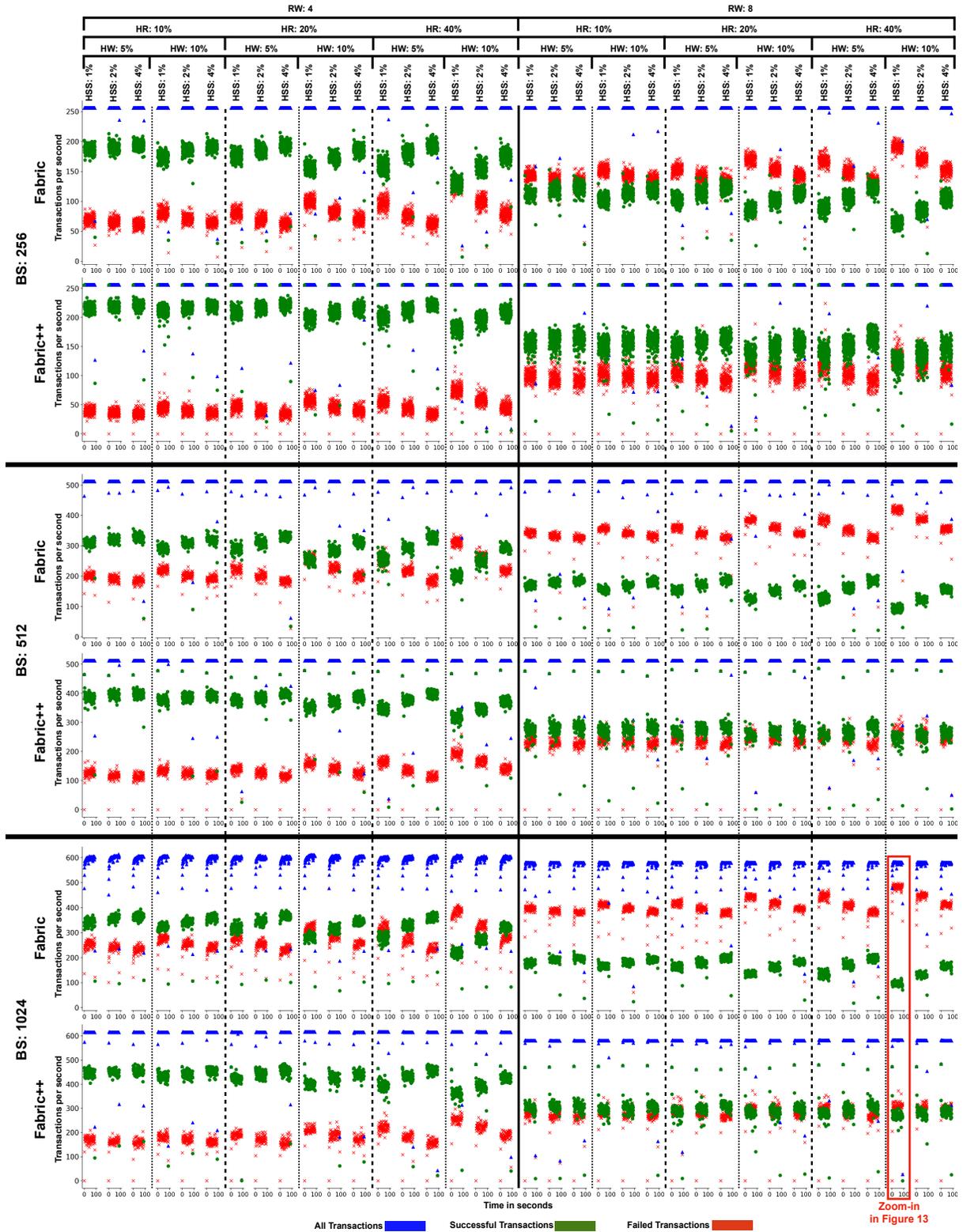}
    \end{center}
    \vspace{-0.5cm}
    \caption{Transactional Throughput of Fabric and Fabric++ under 108 different configurations.}
    \label{figs:configs}
    \vspace{-0.3cm}
\end{figure*} 

We start our experimental evaluation by testing Fabric and Fabric++ under probably the most important criterium for a transaction processing system, namely the throughput of transactions. We differentiate between successful and failed transactions: a good system should try to maximize the number of successful transactions while keeping the number of failed transactions as small as possible.

\begin{table}[!htb]
%\vspace{-0.4cm}
	\caption{Our 108 configurations for the evaluation of the transactional throughput as shown in Figure~\ref{figs:configs}.}
	%\vspace{-0.4cm}
	%\scriptsize
	\setlength{\tabcolsep}{3pt}
	\begin{center}
		\begin{tabular}{ L{12cm} || L{3.3cm} }
			\hline
			\textbf{Experiment Parameters} & \textbf{Values}\\
			\hline
			Fired transaction proposals per second per client & 512\\\hline
			Duration in which transaction proposals are fired & 90 sec\\\hline
			Number of channels & 1\\\hline
			Number of clients per channel & 4 \\
			\hline
			 \multicolumn{2}{c}{} \\
			\hline
			\textbf{System Parameters} & \textbf{Values}\\
			\hline	
			Maximum time to form a block & 1 sec \\\hline
			Maximum number of keys accessed per block& 16384 \\\hline
			Maximum size per block & 2MB \\\hline
			Maximum number of transactions per block (BS) & 256, 512, 1024 \\
			\hline
			\multicolumn{2}{c}{} \\
			\hline
			\textbf{Workload Parameters} & \textbf{Values}\\
			\hline			
			Number of account balances & 10000 \\\hline
			Number of read \& written balances per transaction (RW) & 4, 8\\\hline
			Probability for picking a hot account for reading (HR) & 10\%, 20\%, 40\% \\\hline
			Probability for picking a hot account for writing (HW) & 5\%, 10\% \\\hline
			Number of hot account balances (HSS) & 1\%, 2\%, 4\% \\
			\hline
		\end{tabular}
		\label{table:config}
	\end{center}
	%\vspace{-0.5cm}
\end{table}

To measure this property, we fire a constant stream of transaction proposals for $90$~seconds into a single channel using four clients. Each client fires at a rate of~$512$ proposals per second. This firing rate is sufficient to fully sustain the system in our setup. Table~\ref{table:config} shows the detailed configuration. To identify their impact on the throughput, we vary five important parameters: the maximum number of transactions per block~(BS), the number of read balances and written balances per transaction~(RW), the probability for picking a hot account for reading~(HR) respectively for writing~(HW), as well as the number of hot account balances~(HSS). In total, we evaluate $108$~different configurations in this experiment. 

Figure~\ref{figs:configs} shows the results. First and foremost, we vary the maximum number of transaction per block~(BS), as it has a large impact on the transaction processing in general and the ordering in particular. The results for Fabric and Fabric++ for BS=256 are presented in first and the second row, for BS=512 in the third and fourth row, and for BS=1024 in the fifth and sixth row, respectively. Along the columns, we vary the remaining four parameters RW, HR, HW, and HSS in a total of 36 configurations. For a single run, we show the transactional throughput (\textcolor{blue}{blue}) that was achieved for each second of the $90$~second run. This throughput is additionally split into successful transactions (\textcolor{darkgreen}{green}) and failed transactions (\textcolor{red}{red}). 

To interpret the results, let us look at Figure~\ref{figs:configs} as a whole. We can see that Fabric++ significantly increases the throughput of successful transactions over Fabric for essentially all tested configurations. For vanilla Fabric, we can observe that under configurations accessing many accounts (RW=8), the number of failed transactions per second is actually significantly higher than the throughput of successful transactions. This problem is highly reduced by Fabric++, where the successful throughput is at least on par with the failed throughput, or even dominates it.  The largest improvement of Fabric++ over Fabric in terms of successful transactions we observe is around factor 3x for the configuration BS=1024, RW=8, HR=40\%, HW=10\%, HSS=1\%, which we also show in a zoomed-in version in Figure~\ref{figs:zoomin} of the appendix.
We also observe a significant decrease in the throughput of the successful transactions with the increase in the hotness of the transactions. For large block-sizes (BS $\in \{512, 1024\}$), each block ($b_i$) roughly updates every key in the hotset and a large fraction of coldset. This forces most of the transactions in block ($b_{i+1}$) to abort because of read-write conflicts. So, we observe a pattern of blocks committing with alternating highly-successful and highly-failed transactions. In Fabric, most of the transactions are aborted due to this inter-block conflicts. In addition to this, due to a large block size, Fabric creates a large amount of within-block conflicts, which results in a large fraction of the total number of processed transactions to abort. In Fabric++, we observe a similar alternating behavior in terms of cross-block conflicts. However, since Fabric++ reorders the transactions within the block to remove the within-block conflicts, the number of successful transactions remain on-par with the number of failed transactions. We observe that the strength of Fabric++ lies in contended workload, where the hotness has temporal behavior. If, due to temporal behavior, hot reads, and updates end up in a same block, Fabric++ can possibly optimize the order of transactions to extract a largest set of transactions that have a chance to commit. In contrast to Fabric++, Fabric will behave similarly for temporal and non-temporal hotness in the workloads, forcing a large fraction of transactions to abort, even though they could commit.

Apart from the overall comparison of Fabric and Fabric++, we can analyze the influence of the parameters on the system. A larger block size generally results in a higher throughput. In the case of Fabric++, a larger block size also increases the reordering possibilities of our mechanism. Besides, we can see that a higher number of accesses per transactions results in more failed transactions.

\subsection{Optimization Breakdown}
\vspace{-0.1cm}

\begin{figure}[!htb]
    \begin{center}
        \includegraphics[width=8.5cm, trim={0 0 0 0}, clip]{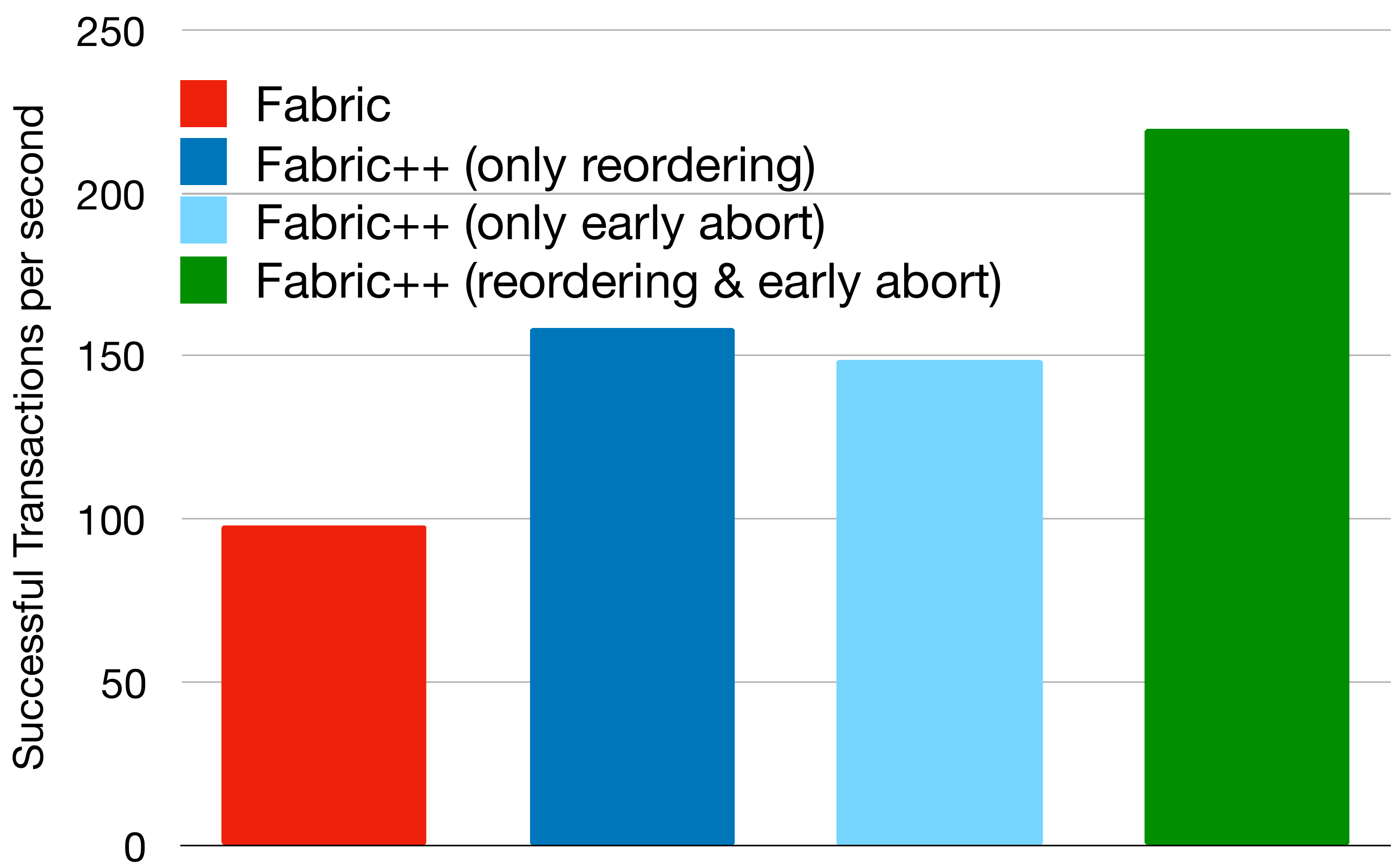}
    \end{center}
    \caption{Breakdown of the individual impact of our optimizations on the throughput of successful transactions for the configuration BS=1024, RW=8, HR=40\%, HW=10\%, HSS=1\%.}
    \label{figs:breakdown}
\end{figure}

In Section~\ref{ssec:throughput}, we measured the throughput of Fabric++ with both optimizations activated. Let us now see at a sample configuration, how much the individual optimizations of reordering and early abort contribute to the improvement. 
Figure~\ref{figs:breakdown} shows the improvement breakdown for the configuration BS=1024, RW=8, HR=40\%, HW=10\%, HSS=1\% in comparison to standard Fabric. While Fabric achieves only a throughput of around $100$~successful transactions per second, activating one of our two optimization techniques alone improves this to around $150$~transactions per second. In comparison to that, activating both techniques at the same time results in the highest throughput of successful transactions with around $220$ transactions per second. This shows nicely how both techniques work together: Transactions, that are already early aborted in the simulation phase do not end up in a block in the ordering phase. As a consequence, only transactions, that have a realistic chance of being successful, are considered in the reordering process. 

\subsection{Scaling Channels and Clients}

So far, in all experiments we used four clients to fire transactions into a single channel. Let us now vary the number of channels as well as the number of clients per channel to see the effect on the throughput. Again, we use the configuration BS=1024, RW=8, HR=40\%, HW=10\%, HSS=1\% and evaluate the average throughput of successful transactions for Fabric and Fabric++.

\begin{figure}[!htb]
%\vspace{-0.3cm}
\newcommand{\swapstretch}{5.0cm}
\subfigure[\textbf{Varying the number of channels} from 1 to 8. Per channel, we use 2 clients to fire the transaction proposals.]{
\includegraphics[width=7.5cm, trim={0 0 0 0}, clip]{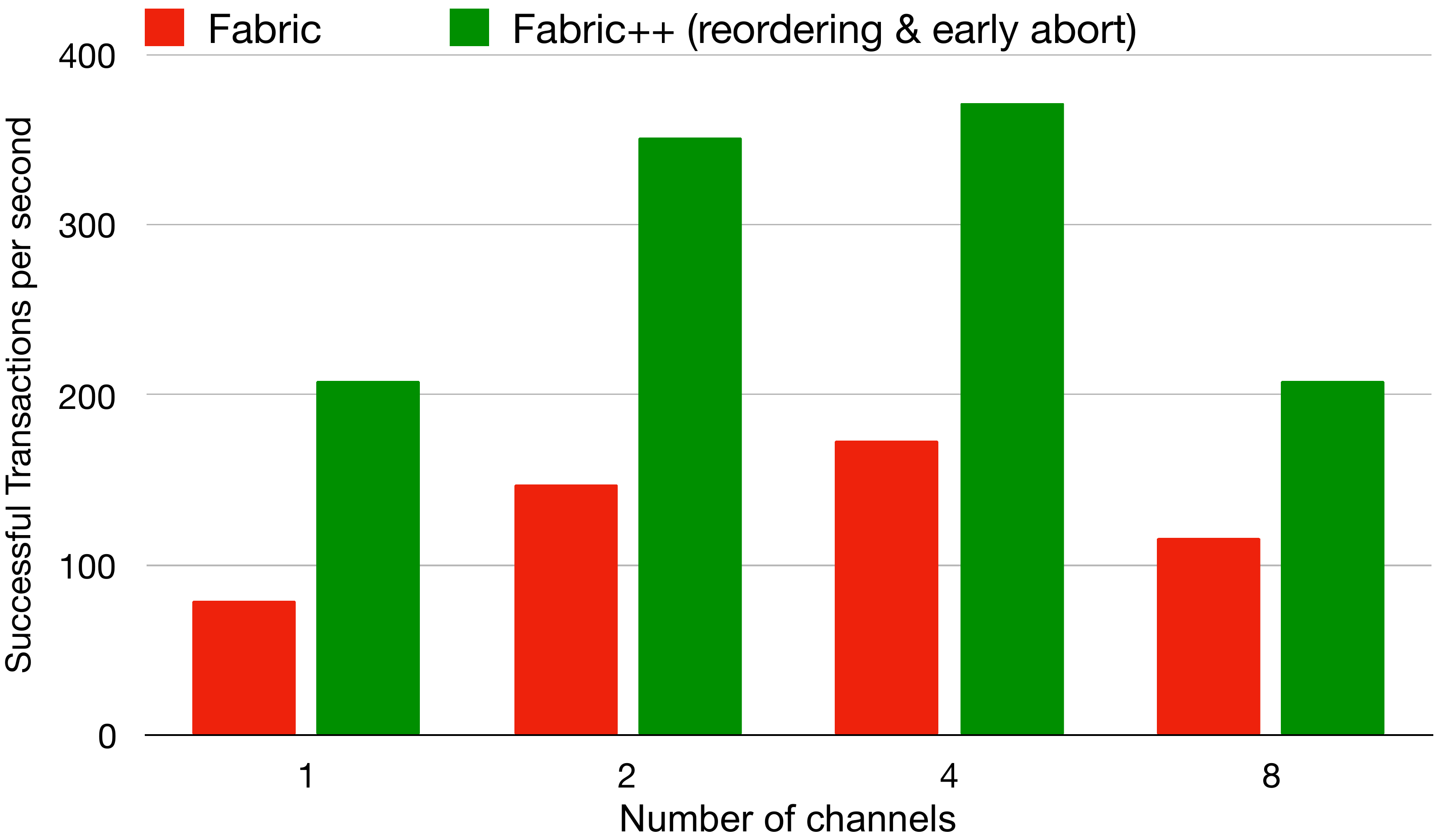}
\label{figs:scaling_channels}
}
\vspace{-0.1cm}
\subfigure[\textbf{Varying the number of clients per channel} from 1 to 8. All clients fire their transaction proposals in a single channel.]{
\includegraphics[width=7.5cm, trim={0 0 0 0}, clip]{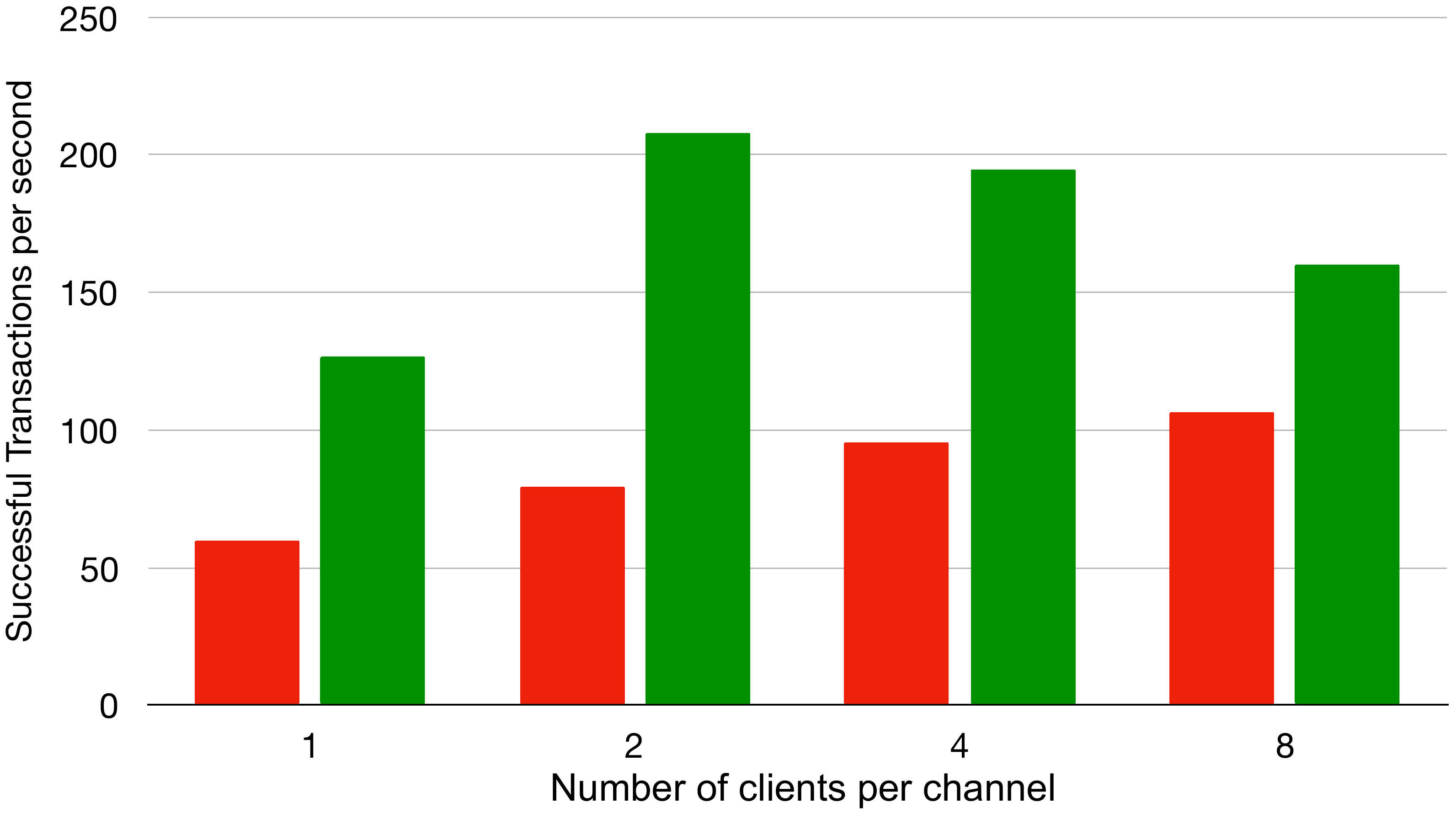}
\label{figs:scaling_clients}
}
%\vspace{-0.2cm}
\caption{The impact of the number of channels as well as the number of clients per channel on the throughput of successful transactions for the configuration BS=1024, RW=8, HR=40\%, HW=10\%, HSS=1\%.}
%\vspace{-0.6cm}
\label{figs:scaling}
\end{figure}

First, we vary the number of channels in Figure~\ref{figs:scaling_channels} from $1$ to $8$. Per channel, we use $2$~clients to fire transaction proposals. We can see that when going from $1$~channel to $4$~channels, the throughput of both Fabric and Fabric++ significantly increases. Obviously, the additional mechanisms of Fabric++ do not harm the scaling with the number of channels. Only when using $8$~channels, the throughput decreases again for both Fabric and Fabric++. This is simply the case because individual channels start competing for resources. This also increases the number of failed transactions: Scaling from $1$ to $8$~channels increases the number of failed transactions from $213$~TPS to $837$~TPS for Fabric and from $81$~TPS to $704$~TPS for Fabric++. Due to the competition for resources, individual simulations phase take longer and increase the chance of working on stale data. 

After varying the number of channels, let us now vary the number of clients per channel in Figure~\ref{figs:scaling_clients}. We test $1$, $2$, $4$, and $8$~clients, where all clients fire their transaction proposals into a single channel. Here, the picture is a slightly different to the behavior when scaling channels. The throughput of Fabric increases very gently with the number of clients, and we see an improvement from around $60$ to $105$ successful transactions per seconds when going from $1$ to $8$~clients. For Fabric++, we see the highest throughput with around $205$ successful transactions per second already for $2$~clients. For $8$~clients, the throughput drops by around factor~$2$ to the throughput of Fabric, clearly showing that the firing clients also compete for resources. This is also visible in an increase in failed transactions when going from $1$ to $8$~clients per channel, which increase from $86$~TPS to $928$~TPS for Fabric and from $20$~TPS to $841$~TPS for Fabric++.

\vspace{-0.3cm}
\section{Conclusion}
\label{sec:conclusion}
\vspace{-0.1cm}
In this work, we identified strong similarities of the transaction pipeline of contemporary blockchain systems at the case of Hyperledger Fabric and distributed database systems in general. We analyzed these similarities in detail and exploited them to transition mature techniques from the context of database systems to Fabric, namely transaction reordering to remove serialization conflicts as well as early abort of transactions, that have no chance to commit. In an extended experimental evaluation, where we tested $108$~different configurations of workload and system, we showed that this improved version Fabric++ outperforms the vanilla Fabric in terms of throughput of successful transactions by up to factor~$3$x, while keeping the scaling capabilities intact.

\subsection*{Acknowledgments }

We would like to thank Immanuel Haffner for helping us in setting up the Fabric cluster, running benchmarks, as well as profiling the internals of the system. 

\bibliographystyle{abbrv}
\bibliography{bibliography_sigmod}

\appendix

%\vspace{-0.3cm}
\section{Throughput Timeline}
\begin{figure}[!htb]
%\vspace{-0.4cm}
\newcommand{\swapstretch}{5.0cm}
\subfigure[Fabric]{
\includegraphics[width=7.5cm, trim={0 0 0 0}, clip]{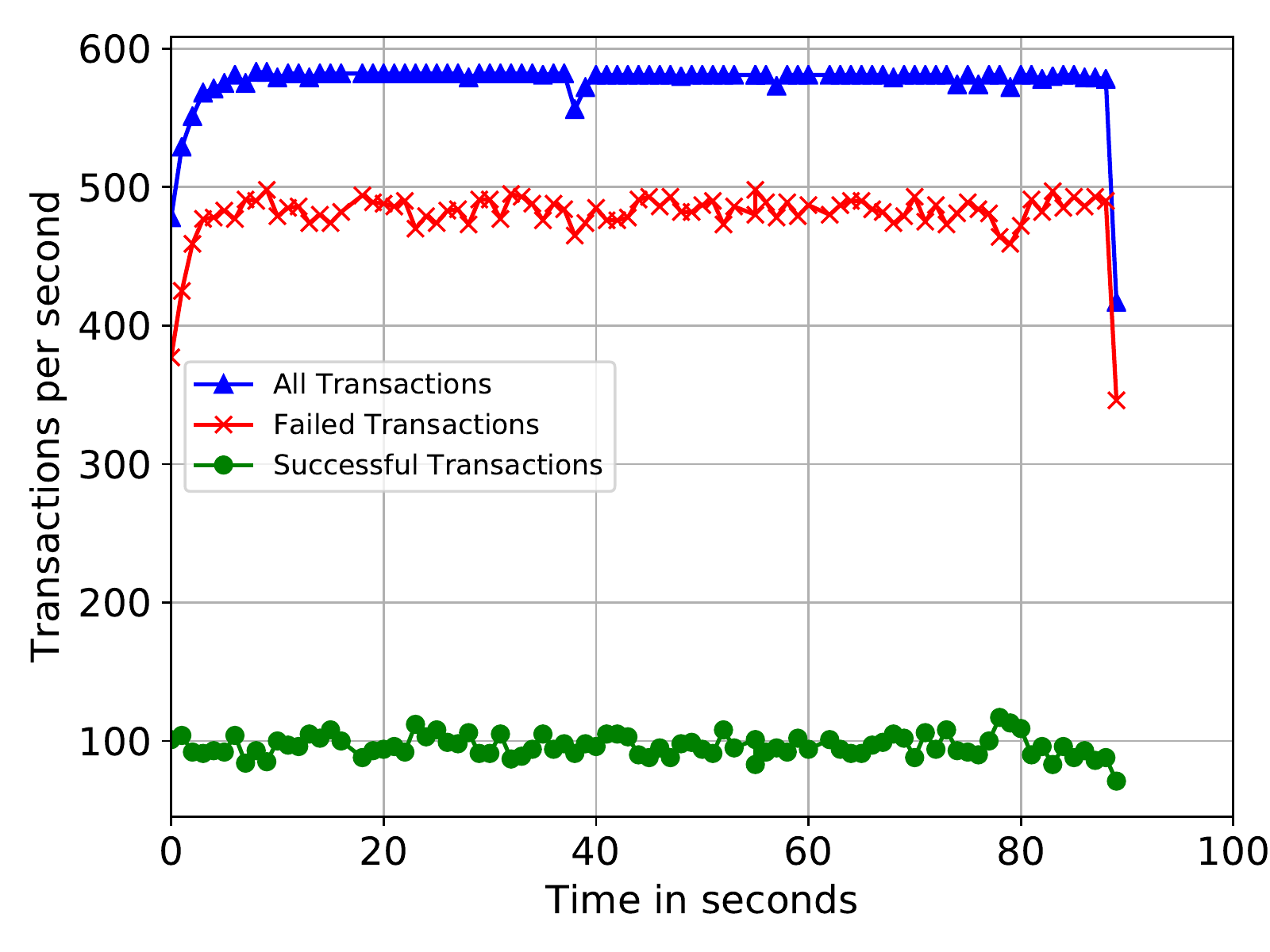}
\label{figs:zoomin_baseline}
}
%\vspace{-0.1cm}
\subfigure[Fabric++]{
\includegraphics[width=7.5cm, trim={0 0 0 0}, clip]{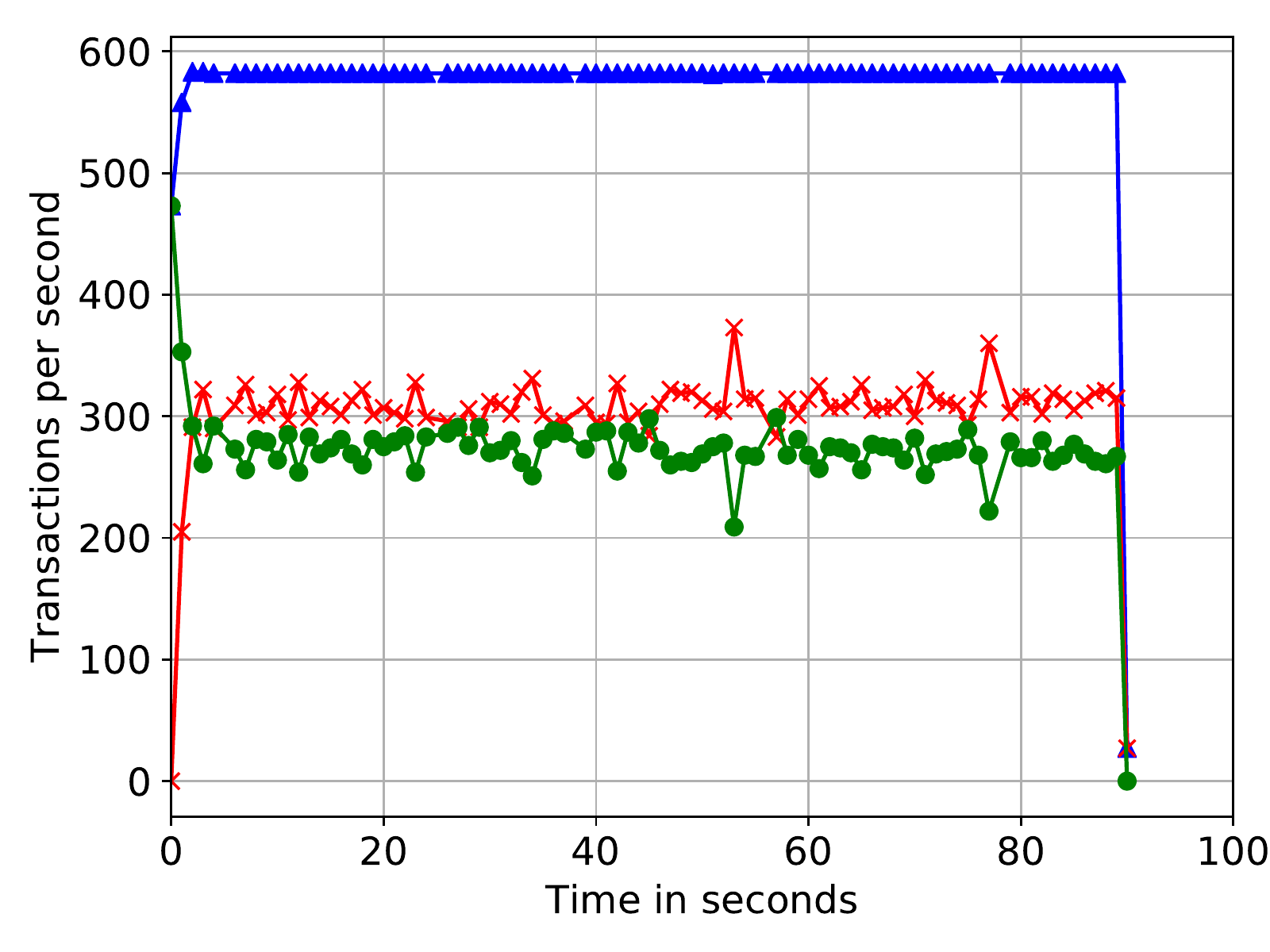}
\label{figs:zoomin_master}
}
%\vspace{-0.2cm}
\caption{Zoom-in of Figure~\ref{figs:configs} for the configuration BS=1024, RW=8, HR=40\%, HW=10\%, HSS=1\%.}
%\vspace{-0.2cm}
\label{figs:zoomin}
\end{figure}

%\vspace{-0.1cm}
Figure~\ref{figs:zoomin} presents the detailed zoom-in of the run for configuration BS=1024, RW=8, HR=40\%, HW=10\%, HSS=1\% for Fabric (Figure~\ref{figs:zoomin_baseline}) and Fabric++ (Figure~\ref{figs:zoomin_master}). We can see that the throughput remains very stable over the run of $90$~seconds. In the beginning, there is a small ramp-up phase visible, which is actually very interesting. For Fabric, the throughput of successful transactions directly starts very low with only $100$ transactions per second. In contrast to that, for Fabric++, the initial throughput of successful transactions almost reaches $500$~transactions per second with the number of failed transactions per second being $0$. This shows that for the first block, our reordering mechanism manages to completely resolve all intra-block conflicts. After that, inter-block conflicts can arise which increase the number of failed transactions in any case.   

%\vspace{-0.3cm}
\section{Ordering Service Micro-Benchmark 2: Vary the length of cycles}
\label{sec:ordering_benchmark2}
\begin{figure}[!htb]
    \vspace{-0.2cm}
    \begin{center}
        \includegraphics[width=8.5cm]{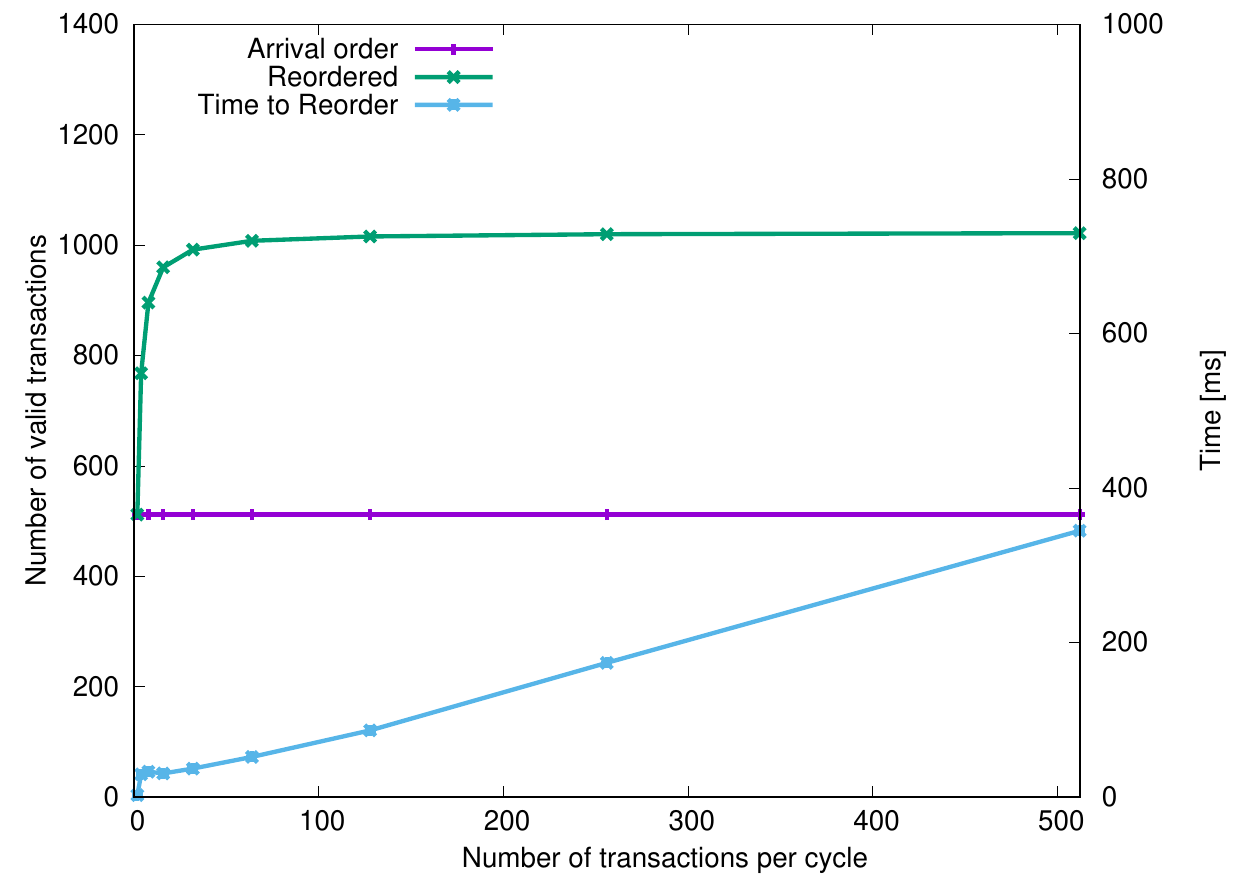}
    \end{center}
    \vspace{-0.5cm}
    \caption{Workload 2: Varying the length of the cycles.}
    \label{figs:ordering_benchmark2}
    \vspace{-0.3cm}
\end{figure}

%\vspace{-0.1cm}
In the following experiment, we want to analyze the impact of cycles on the arrival order and on our reordering mechanism. To do so, we again form a sequence of $n$~transactions, that contains $n/t$~cycles of size~$t$ transactions of the form 
\begin{small}
$$ T[\textcolor{blue}{r(k_0)},  \textcolor{red}{w(k_0)}], T[\textcolor{blue}{r(k_0)},  \textcolor{red}{w(k_1)}], T[\textcolor{blue}{r(k_1)},  \textcolor{red}{w(k_2)}], T[\textcolor{blue}{r(k_2)},  \textcolor{red}{w(k_0)}] $$ 
\end{small}
Again, we want to identify how many transactions are valid under the arrival order and when using our reordering mechanism. Figure~\ref{figs:ordering_benchmark2} shows the results for $1024$~transactions. For the arrival order, only half of transactions are valid, no matter of the cycle length. This is because aborting every second transaction breaks the cycles. In comparison to that, our reordering mechanism is able to achieve a high number of valid transactions, if the cycles are sufficiently long respectively, there are not too many cycles to cancel. Of course, our algorithm becomes more expensive with the length of the cycles to break. However, since extremely long cycles are very unlikely to occur in reality, the runtime of our mechanism will in general remain low in the ordering phase, as we will see in the full fledged evaluation later on. 

\section{Extended Throughput Evaluation}
\label{sec:extended_throughput}
Additionally to 10,000 account balances, as used in the previous experimental evaluation, we test Fabric and Fabric++ as well under 20,000 account balances and 2 read and write accesses per transaction (RW=2). We can observe that for RW=2, the number of successful transactions is significantly higher than the number of failed transactions due to less conflicts than for RW=4 or RW=8. 

\begin{figure*}[!htb]
    \vspace{-0.2cm}
    \begin{center}
        \includegraphics[width=\textwidth, trim={0 0 0 0}, clip]{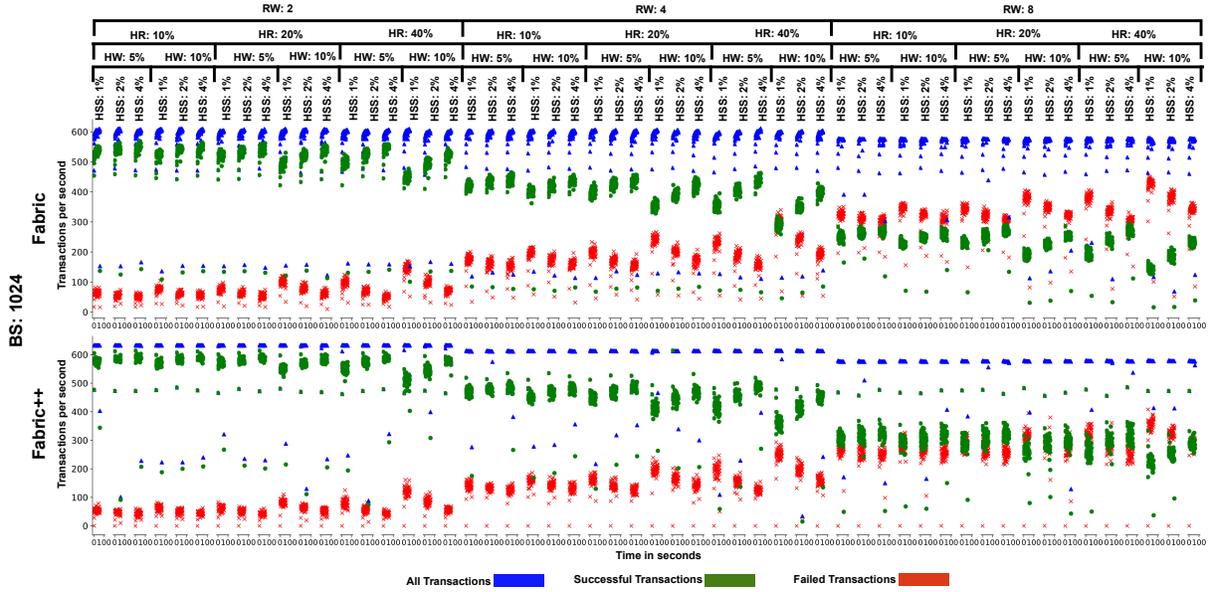}
    \end{center}
    \vspace{-0.5cm}
    \caption{Transactional Throughput of Fabric and Fabric++ under 54 different configurations, where the number of account balances is 20,000.}
    \label{figs:configs_20k}
    \vspace{-0.3cm}
\end{figure*}

\end{document}